\documentclass[fleqn,usenatbib]{mnras}

\usepackage{newtxtext,newtxmath}
\usepackage{amsmath}
\usepackage[T1]{fontenc}

\DeclareRobustCommand{\VAN}[3]{#2}
\let\VANthebibliography\thebibliography
\def\thebibliography{\DeclareRobustCommand{\VAN}[3]{##3}\VANthebibliography}

\usepackage{graphicx}	
\usepackage{amsmath}	

\usepackage{lineno}

\title[Modelling stars with GP]{Modelling stars with Gaussian Process Regression: Augmenting Stellar Model Grid}

\author[T. Li et al.]{
Tanda Li$^{1}$\thanks{E-mail: t.li.2@bham.ac.uk},
Guy R. Davies$^{1}$\thanks{E-mail: G.R.Davies@bham.ac.uk},
Alexander J. Lyttle$^{1}$,
Warrick H. Ball$^{1}$,
Lindsey M. Carboneau$^{1}$,
\newauthor
Rafael A. Garc\'ia$^{2}$,
\\
$^{1}$ School of Physics and Astronomy, University of Birmingham, Birmingham, B15 2TT, United Kingdom\\
$^{2}$ AIM, CEA, CNRS, Universit\'e Paris-Saclay, Universit\'e Paris Diderot, Sorbonne Paris Cit\'e, F-91191 Gif-sur-Yvette, France
}

\date{Accepted XXX. Received YYY; in original form ZZZ}

\pubyear{2020}

\begin{document}
\label{firstpage}
\pagerange{\pageref{firstpage}--\pageref{lastpage}}
\maketitle

\begin{abstract}
Grid-based modelling is widely used for estimating stellar parameters. However, stellar model grid is sparse because of the computational cost. This paper demonstrates an application of a machine-learning algorithm using the Gaussian Process (GP) Regression that turns a sparse model grid onto a continuous function. We train GP models to map five fundamental inputs (mass, equivalent evolutionary phase, initial metallicity, initial helium fraction, and the mixing-length parameter) to observable outputs (effective temperature, surface gravity, radius, surface metallicity, and stellar age). 
%
%
We test the GP predictions for the five outputs using off-grid stellar models and find no obvious systematic offsets, indicating good accuracy in predictions. 
As a further validation, we apply these GP models to characterise 1,000 fake stars. Inferred masses and ages determined with GP models well recover true values within one standard deviation. 
An important consequence of using GP-based interpolation is that stellar ages are more precise than those estimated with the original sparse grid because of the full sampling of fundamental inputs.
\end{abstract}




\section{Introduction}

Theoretical stellar model has been developed for decades to simulate star structure and evolution. Star modelling is mostly grid-based \citep[e.g.][]{2016ApJ...823..102C} because computing many stellar models are time-consuming especially when a number of free input parameters are considered. Varying one of these adjusted parameters (mass, metallicity, helium fraction, mixing-length parameter, etc.) adds on an input dimension and hence exponentially increases the computational cost. 

A sparse grid is not ideal for the statistics analysis. Classical interpolation method has been applied to overcome this disadvantage. For instance, \citet{2016ApJS..222....8D} applied a method to transform stellar evolution tracks onto a uniform basis and then interpolate to construct stellar isochrones. More recently, \citet{2019MNRAS.484..771R} uses Bayesian statistics and a Markov Chain Monte Carlo approach to find a representative set of interpolated models from a grid. Their approaches based on multivariate linear interpolation achieve good accuracy for 3-dimension girds (inputs are mass, age, and metallicity). However, it hasn't been studied in detail for higher-dimension grid. For example, \citet{2021MNRAS.500...54N} interpolated a 5-dimension stellar model grid to study the initial helium abundance, but they did not present an analysis of the interpolation uncertainties. The difficulty in applying the classical interpolation method to the high dimension problem is the choice of the function form.  Stellar tracks do not follow the same scale function through all dimensions and all parameter ranges. For instance, the evolutionary tracks on the HR diagram changes its shape at the mass of $\sim$1.1 ${\rm M_{\odot}}$ because of the switch between radiative and convective core. In this mass range, using the linear interpolation would not give good predictions for stellar models about the blue hook.

Compared with the classical interpolation with specific function form, the machine-learning tools can offers flexible functional forms to handle high-dimension problems. Machine learning is being applied to the field of stellar research in many ways to efficiently characterise stars.
\citet{2016MNRAS.461.4206V} applied artificial neural network, which is a series of algorithms that endeavours to recognise underlying relationships in a set of data, to determine the evolutionary parameters of the sun and sun-like stars based on spectroscopic and seismic measurements. Using a similar artificial neural network interference, \citet{2019PASP..131j8001H} developed a method to provide the optimal starting point of model competitions for more detailed forward asteroseismic modelling. Moreover, \citet{2021arXiv210313394M} trained neural networks to predict theoretical pulsation periods of high-order gravity modes, as well as the luminosity, effective temperature, and surface gravity for a given mass, age, overshooting parameter, diffusive envelope mixing, metallicity, and near-core rotation frequency. 
Using different machine learning tools, \citet{2016ApJ...830...31B} trained a random forest regressors \citep{ho1995random}, which is an ensemble learning method for regression and operates by constructing a multitude of decision trees, to rapidly estimate fundamental parameters of solar-like stars based on classical and asteroseismic observations. \citet{2018MNRAS.476.3233H} developed a convolutional neural network classifier that analyses visual features in asteroseismic frequency spectra to distinguish between red giant branch stars and helium-core burning stars. \citet{2019MNRAS.484.5315W} determined masses and ages for massive RGB stars from their spectra with a machine-learning method based on kernel principal component analysis, which is a nonlinear form of principal component analysis using integral operator kernel functions and can efficiently compute principal components in high dimensional feature spaces related to input space by some nonlinear map \citep{scholkopf1997kernel}. \citet{2020MNRAS.499.2445H} applied the mixture density network \citep{bishop1994mixture}, which learns a transformation from a set of input variables to a set of output variable, to determine stars' fundamental parameters like mass and age based on observed mode frequencies, spectroscopic, and global seismic parameters.

In above studies, the discriminative machine-learning model is mostly used. The discriminative model treats observables as given facts to directly infer star fundamental parameters. The method is efficient and easy for computation, while the downside is not allowing any priors for star properties like mass.
In an opposite direction, the generative machine-learning model uses the star fundamental parameters as given facts to predict observables. This approach offers flexibility to prior fundamental parameters in the sampling. For instance,  \citet{2021MNRAS.tmp.1343L}  determined initial helium fraction and mixing-length parameters for a sample of {\em Kepler} dwarfs and subgiants with an artificial neural network to provide the generative model. This allowed them to prescribe prior distributions over the fundamental stellar parameters and, by extension, over population-level parameters such as a helium enrichment law. Priors encode our current knowledge and assumptions into inference from new data. This is especially important with noisy observations which span a large portion of parameter-space.

Constructing a comprehensive and fine model grid is computationally expensive. In this work, we aim to apply the machine learning tool to transform a sparse model grid onto a continuous function.  We apply a machine learning algorithm that involves a Gaussian process (GP) that measures the similarity between data points (i.e., the kernel function) to predict values for unseen points from training data. We use the generative model and treat fundamental parameters as given facts to predict observables. This gives us flexibility to prior fundamental inputs when modelling stars. 
We organise the rest of the paper as follow. Section \ref{sec:grid} contents descriptions about the computation of a representative stellar model grid. We then introduce the underlying theory of GP and the setup of GP model in Section \ref{sec:gpmodel}. We then demonstrate some preliminary studies for low-dimension problems in Section~\ref{examples}. Section \ref{sec:results} demonstrates GP predictions and their systematic uncertainties. Subsequently, we augment the grid to have a set of continuously-sampled stellar models and model 1,000 fake stars for testing the accuracy of our method in Section \ref{sec:augmentation}. Lastly, we discuss advantages and limitations of this approach, highlight areas where improvements can be found in the near future, and summary conclusions in Section \ref{sec:conclusion}.











\section{Representative Model Grid}\label{sec:grid}

\subsection{Grid computation}

We compute a stellar model grid as the training dataset. We aim to cover stars with approximate solar mass on the main-sequence and the subgiant phases. We consider four independent fundamental inputs which are stellar mass ($M$), initial helium fraction ($Y_{\rm init}$), initial metallicity ([Fe/H]$_{\rm init}$), and the mixing-length parameter ($\alpha_{\rm MLT}$). 
We calculated three model dataset for different purposes. The primary dataset is a standard model grid with uniform mass step. This model grid is used for all preliminary tests and also for the final training. 
We also calculate an additional dataset to increase the grid resolution for $M$ > 1.05$\rm M_{\odot}$, because we find that the blue hook feature (where global parameters sharply vary) is relatively hard to train. This dataset is only used in the final training. 
Details of parameter ranges and steps of the two grids are listed in Table \ref{tab:grid}. 
For validating and testing GP predictions, we computed off-grid models with randomly sampled fundamental inputs as a third dataset. 
%
%
The computation of evolutionary tracks starts at the Hayashi line with pre-main-sequence central temperature at 300,000K and terminates at the base of red-giant branch (RGB) where $\log g$ = 3.6dex. Note that we only use models after the zero-age-main-sequence (ZAMS), which is defined as the point where core-hydrogen burning contributes over 99.9\% of the total luminosity. 

\begin{table}
	\centering
	\caption{Computation of Stellar model grid.}
	\label{tab:grid}
	\begin{tabular}{llll} 
		\hline
		\multicolumn{3}{c}{Primary dataset}\\
		\hline
		Input Parameter & Range & Increment \\
        \hline
	$M$ ($\rm M_{\odot}$) & 0.80 -- 1.20 &  0.01\\
        $\rm{[Fe/H]}$ (dex) & -0.5 -- 0.2/0.2 -- 0.5 & 0.1/0.05\\
        	$Y_{\rm init}$ & 0.24 -- 0.32 & 0.02\\
        $\alpha_{\rm{MLT}}$  & 1.7 -- 2.5&  0.2\\
        \hline
       \multicolumn{3}{c}{Additional dataset}\\
	\hline
	Input Parameter & Range & Increment \\
        \hline
	$M$ ($\rm M_{\odot}$)  & 1.055 -- 1.195 &  0.01\\
        $\rm{[Fe/H]}$ (dex) & 0.25 -- 0.45 & 0.1\\
        	$Y_{\rm init}$ & 0.25 -- 0.31 & 0.02\\
        $\alpha_{\rm{MLT}}$  & 1.8-- 2.4&  0.2\\
	\hline
	\end{tabular}
\end{table}

\subsection{Input physics}\label{subsec:stellar_model}

We use the stellar code Modules for Experiments in Stellar Astrophysics
(\textsc{MESA}, version 12115) to construct stellar grids. 
\textsc{MESA} is an open-source stellar evolution package which is undergoing active development. 
Descriptions of input physics and numerical methods
can be found in \citet{2011ApJS..192....3P,2013ApJS..208....4P, 2015ApJS..220...15P}.
We adopted the solar chemical mixture [$(Z/X)_{\odot}$ = 0.0181]
provided by \citet{2009ARA&A..47..481A}. 
The initial helium fraction ($Y_{\rm init}$) and initial metallicity ($\rm{[Fe/H]_{init}}$) are both independent inputs. 

We use the \textsc{MESA} $\rho-T$ tables based on the 2005
update of OPAL EOS tables \citep{2002ApJ...576.1064R} and OPAL opacity
supplemented by low-temperature opacity \citep{2005ApJ...623..585F}. 
The grey Eddington $T-\tau$ relation is used to determine boundary conditions for modelling the atmosphere.
The mixing-length theory is implemented and the convection is adjusted by the mixing-length parameter ($\alpha_{\rm MLT}$).
We also apply the \textsc{MESA} convective premixing scheme \citep{2019ApJS..243...10P}, which an approach to handling mixing in convection zones that improves model structures at the convective boundary.
Atomic diffusion of helium and heavy elements was also taken into account. MESA calculates particle diffusion and gravitational settling by solving Burger's equations using the method and diffusion coefficients of \citet{Thoul94} as well as  radiative turbulence formula given by \citet{2002A&A...390..611M}.
We consider eight elements (${}^1{\rm H}, {}^3{\rm He}, {}^4{\rm He}, {}^{12}{\rm C}, {}^{14}{\rm N}, {}^{16}{\rm O}, {}^{20}{\rm Ne}$, and ${}^{24}{\rm Mg}$) for diffusion calculations, and have the charge calculated by the MESA ionization module, which estimates the typical ionic charge as a function of $T$, $\rho$, and free electrons per nucleon from \citet{Paquette1986}. We only compute diffusion during the main-sequence stage before the central hydrogen abundance drops below 0.05, because its effects can be neglected in post main-sequence stages. 
The \textsc{MESA} inlist used for the computation is available on \url{https://github.com/litanda/mesa_inlist/}.  



\subsection{Equivalent Evolutionary Phase}

\begin{figure*}
        \includegraphics[width=1.\columnwidth]{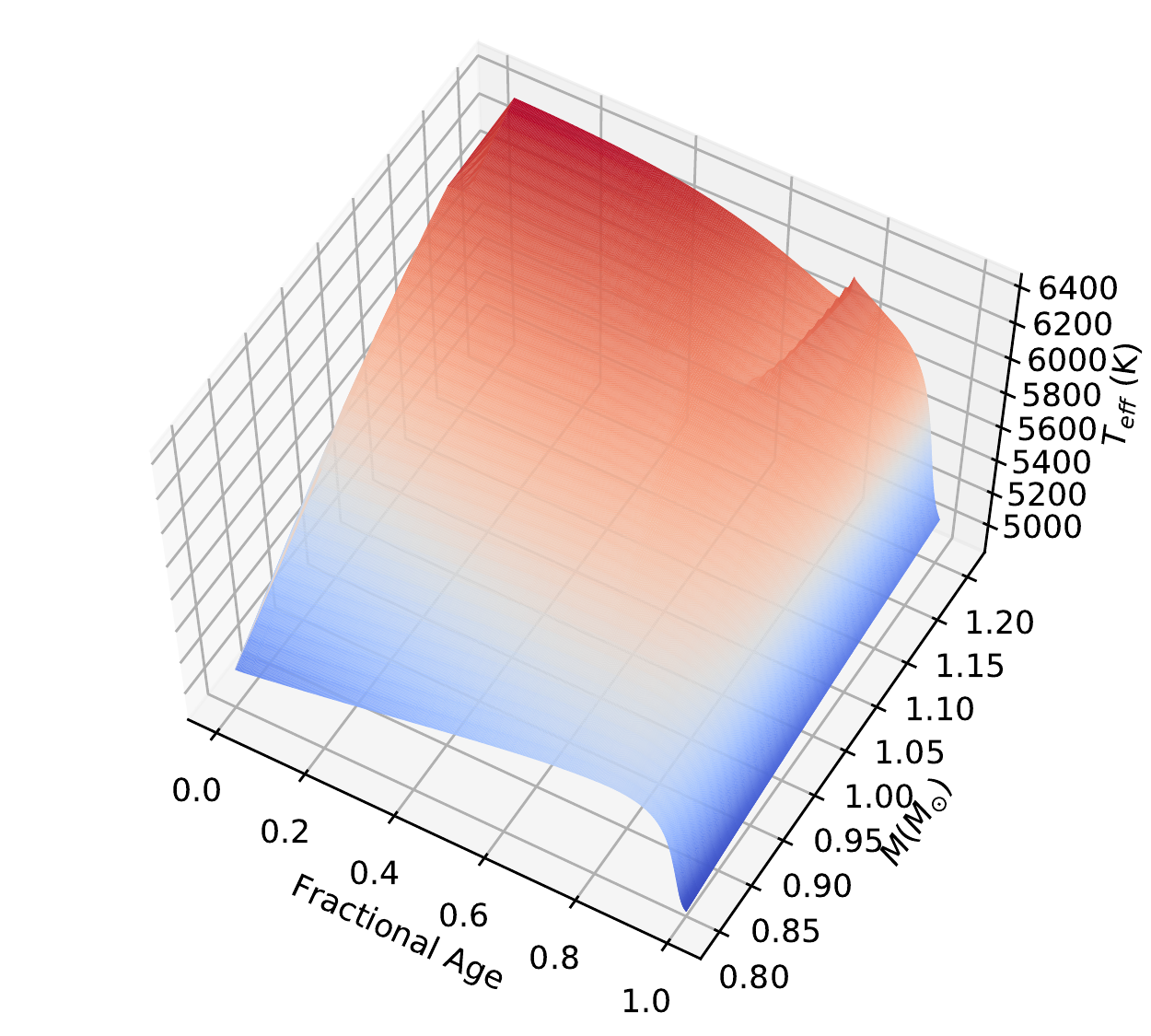}
	\includegraphics[width=1.\columnwidth]{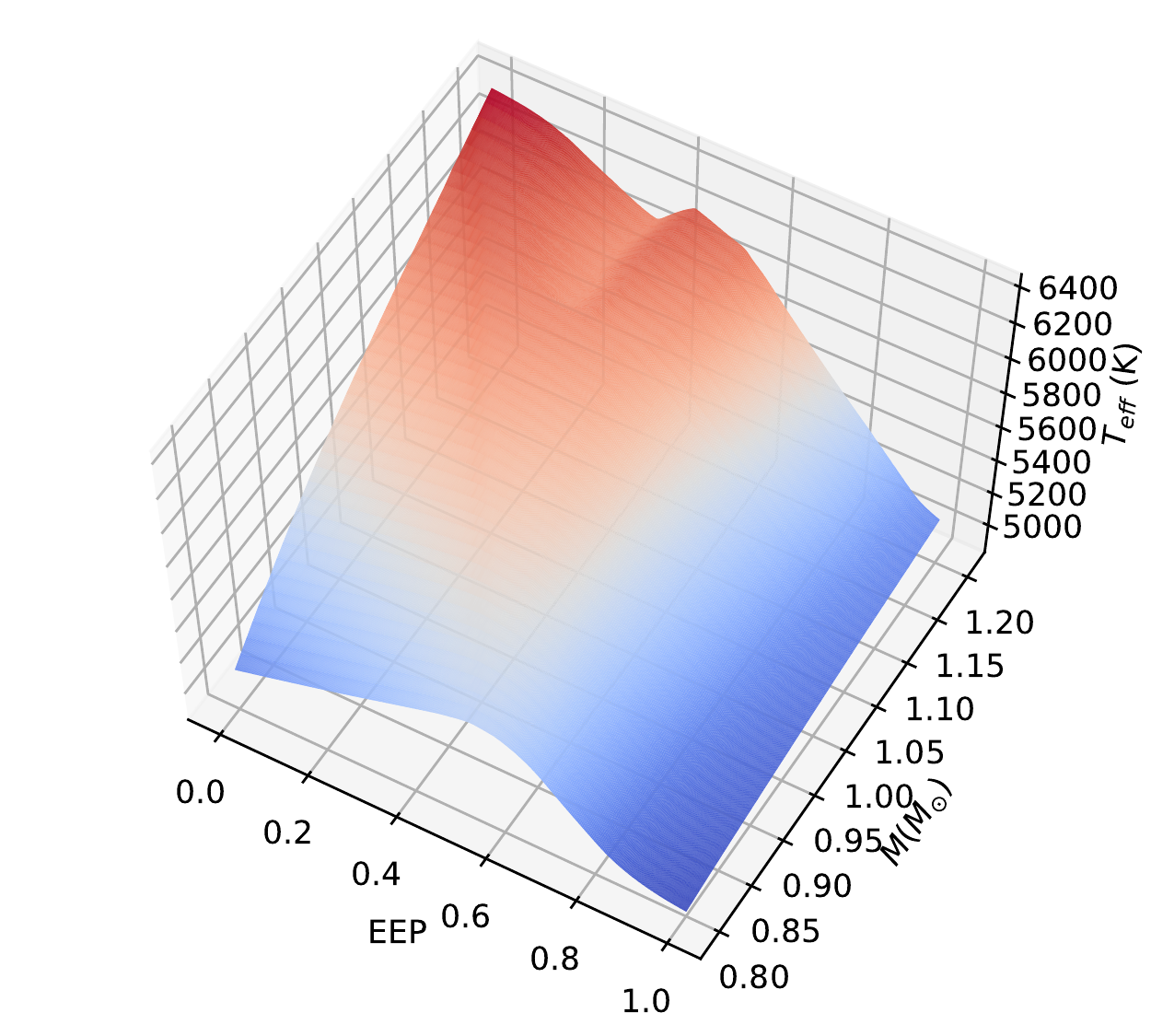}
     \caption{Surface plots of model effective temperature on the mass-fractional age (left) and mass-EEP (right) diagrams. Models in this figure are from the primary grid with fixed initial metallicity ($\rm [Fe/H]_{init}$ = 0.0), helium fraction ($Y_{\rm init}$ = 0.28) and mixing-length parameter ($\alpha_{\rm MLT}$ = 2.1). It can be seen that the effective temperature changes much smoother on the mass-EEP diagram at the blue hook and turn-off points.}
    \label{fig:eep}
\end{figure*}

Apart from the four independent model inputs, i.e., mass, metallicity, helium fraction, and the mixing-length parameter, stellar age is the fifth fundamental input. However, the dynamical range of age varies track by track. This makes GP models hard to map from age to global parameters. 
We need a uniform input to replace the age. The fractional age is an option but we find that global parameters (e.g. effective temperature) sharply change with the fractional age around the blue hook and the turn-off point (as shown in the left panel in Figure~\ref{fig:eep}). It requires a complex and spiky kernel function to fit the curvatures in this area and hence difficult for GP to learn. A quantity Equivalent Evolutionary Phase ({\it EEP}) has been introduced in some model databases like BaSTI, PARSEC, and MIST \citep{2012MNRAS.427..127B,2016ApJS..222....8D, 2018ApJ...856..125H}. The {\it EEP} numbers evolutionary stages and transform stellar tracks for different masses onto a uniform basis.
We follow this idea but define {\it EEP} in a different way to make global parameters change relatively smoothly.
On each evolutionary track, we compute the displacement between consecutive models on the $\log T_{\rm eff} - \log g$ diagram. For instance, the displacement between model $n$ and model $n-1$ can be calculated as
\begin{equation}\label{eq:disp}
\delta d_{n} = ((\log T_{\rm eff, n} - \log T_{\rm eff, n-1}) ^{2} + (\log g _{n} - \log g_{n-1})^{2})^{c},
\end{equation}
where $c$ is an adjusted parameter to scale the displacement. 
The total displacement of model $n$ from the ZAMS (model 0) can be calculated with
\begin{equation}
d_{n} = \sum_{i = 1}^{i = n} \delta d_{i} .
\end{equation}
We then normalise $d_{n}$ to the 0 --1 range and define it as {\it EEP}. On an evolutionary track, {\it EEP} equals to 0 at the ZAMS and 1 on the RGB where $\log$ = 3.6dex. The factor $c$ in Eq. \ref{eq:disp} is introduced for modulating {\it EEP} because the track step on the $\log T_{\rm eff} - \log g$ diagram is not uniform. To avoid obvious data gaps, we test some cases and find that $c$ = 0.18 gives the most uniform data distribution.
In Figure~\ref{fig:eep}, we demonstrate how the effective temperature changes with fractional age and {\it EEP}. It can be seen that {\it EEP} is a better choice than the fractional age because global parameters change smoother around the blue hook and the turn-off point.

\subsection{Sampling method}\label{sec:selection}

%
There is a limitation of the data size in the GP framework, because the computational and memory complexity exponentially increase with the number of data points. In practice, the typical data size is on an order of $10^{4}$. Given that the grid contains $\sim 10,000,000$ stellar models, only a small subset can be used for training. The sampling method is hence important.
 A flat sampling is not appropriate, because the evolving step is not uniform at different evolutionary stages due to the \textsc{MESA} step-control strategy. For instance, stellar models are dense at the main-sequence and lower RGB but quite sparse at the subgiant stage. We test a few methods and find that using the displacement ($\delta d_{n}$) defined in Eq.~\ref{eq:disp} as the weight to sample models on an evolutionary track gives a relatively uniform data distribution at different evolutionary stages. 

\section{Gaussian Process Model}\label{sec:gpmodel}
A GP can be applied as a probabilistic model to a regression problem.  Here we use the GP model to generalise a stellar model grid to a continuous and probabilistic function that maps inputs to observable quantities.  This allows us to predict observable quantities for off-grid regions.
We intend to train GP models that maps five fundamental inputs,  i.e., mass ($M$), initial metallicity ([Fe/H]$_{\rm init}$), initial helium fraction ($Y_{\rm init}$), the mixing-length parameter ($\alpha_{\rm MLT}$), and equivalent evolutionary phase ({\it EEP}), to five model outputs including effective temperature ($T_{\rm eff}$), surface gravity ($\log g$), radius ($R$), surface metallicity ({[Fe/H]}), and stellar age ($\tau$). 
We only choose three global parameters ($T_{\rm eff}$, $\log g$, $R$) as GP outputs because this work is mainly for demonstrating and testing the method. Other global parameters like the luminosity and the seismic large separation can also be trained and predicted. However, because many global quantities are correlated to others, examining consistences of their predictions are necessary. 
We use the GP model as a non-parametric emulator, that is emulating the comparatively slow calls to models of stellar evolution.
This emulator can be described as a function approximation problem. In fact, the way we have implemented the GP as function approximation means that we have used one GP for each of the outputs so that they can be described as
\begin{equation}\label{gprmodel1}
{T_{\rm eff}} = f_{T_{\rm eff}}(M, EEP, {\rm [Fe/H]}_{\rm init}, Y_{\rm init}, \alpha_{\rm MLT}),
\end{equation}
\begin{equation}\label{gprmodel1}
{\log g} = f_{\log g}(M, EEP, {\rm [Fe/H]}_{\rm init}, Y_{\rm init}, \alpha_{\rm MLT}),
\end{equation}
\begin{equation}\label{gprmodel1}
{R} = f_{R}(M, EEP, {\rm [Fe/H]}_{\rm init}, Y_{\rm init}, \alpha_{\rm MLT}),
\end{equation}
\begin{equation}\label{gprmodel1}
{\rm [Fe/H]} = f_{\rm [Fe/H]}(M, EEP, {\rm [Fe/H]}_{\rm init}, Y_{\rm init}, \alpha_{\rm MLT}),
\end{equation}
and 
\begin{equation}\label{gprmodel1}
{\tau} = f_{\tau}(M, EEP, {\rm [Fe/H]}_{\rm init}, Y_{\rm init}, \alpha_{\rm MLT}).
\end{equation}
In the following, we introduce the underlying theory of GP regression and the setup of training GP models.

\subsection{Gaussian Process Application}

In our application to a stellar model grid, a GP has a number of desirable properties. While a GP is a stochastic process, the distribution of a GP can be considered as a distribution of functions with a continuous domain.  In fact,  the marginal likelihood considered in function space is equal to the likelihood of the data given some function values,  multiplied by the prior on those function values marginalised over all function values \citep{williams1996gaussian}.  That is to say that, the GP allows for the analytical evaluation of a fit over many different functions (perhaps an infinite number) weighted by some concept of a prior and the agreement with the data. In addition, while the marginal likelihood will be assessed on discrete data,  predictions can be made using linear algebra for new data in the continuous domain, but crucially again marginalised over these many different functional forms.  It is possible to see how this might be useful for generalising (or emulating or augmenting) a discrete grid of stellar models in order to obtain predictions in the continuous domain.

In this section we will look at the required mathematics to be able to implement a GP for our application to grids of stellar models.  We start with a series of definitions before dealing with the marginal likelihood and the posterior predictive distributions. 

We start with a grid of stellar models containing $N$ models with a label we want to learn, for example model effective temperature, which we will denote with the general symbol $\bf y$, and a set of on-grid inputs $\bf X$ (e.g., mass,  {\it EEP},  metallicity,  ...).  We can use a GP to make predictions of the effective temperature (labelled $y$) for additional off-grid input values given by $\bf X_{\star}$.  The vector $\bf y$ is arranged ${\bf y} = \left(y_{i}, ... ,y_{N} \right)^{T}$ where the subscript label references the stellar model.  The input labels are arranged into a $N \times D$ matrix where $D$ is the number of input dimensions (e.g., $D=3$ for mass, {\it EEP}, and metallicity) so that ${\bf X} = ({\bf x}_{1}, ..., {\bf x}_{N})^{T}$ where ${\bf x_{i}} = (x_{1, i}, ..., x_{D, i})^{T}$.  The matrix of additional inputs $\bf X_{\star}$ has the same form as $\bf X$ but size $N_{\star} \times D$.

\citet{williams1996gaussian}, from which our description below is based,  define a GP as a collection of random variables, where any finite number of which have a joint Gaussian distribution.  In general terms,  a GP may be written so that our on grid labels are random variables drawn from our GP distribution, 
\begin{equation}
{\bf y}({\bf X}) \sim \mathcal{GP}\left( m({\bf X}),  {\bf \Sigma}\right),
\end{equation}
where $m({\bf X})$ is some mean function, and $\bf \Sigma$ is some covariance matrix.  The mean function controls the deterministic part of the regression and the covariance function controls the stochastic part.  The mean function defined here could be any deterministic function and we will label the additional parameters, or hyperparameters, $\phi$.  Each element of the more familiar covariance matrix is defined by the covariance function or {\it kernel function} $\bf K$ which has hyperparameters $\theta$ and is given by,
\begin{equation}
{\bf \Sigma} = {\bf K}({\bf X}, {\bf X},  \theta),
\end{equation}
or 
\begin{equation}
{\bf \Sigma}_{n, m} = k({\bf X}_{n}, {\bf X}_{m},  \theta),
\end{equation}
where the inputs ${\bf X}_{n}$ and ${\bf X}_{n}$ are $D$-dimensional vectors and the output is a scalar covariance.
In addition to the covariance defined by the kernel function, we include additional white noise in the covariance matrix by adding an identity matrix $\mathcal{I}$ multiplied by a scalar value $\sigma_{w}^2$, so that, 
\begin{equation}
{\bf \Sigma} = {\bf K}({\bf X}, {\bf X},  \theta) + \sigma_{w}^{2} \mathcal{I},
\end{equation}
where $\sigma_{w}^2$ is another hyperparameter to be learnt during training. 

\subsubsection{The likelihood}
Conceptually we value the GP because of it's ability to marginalise over many functions $\bf f$ and return a marginal likelihood,
\begin{equation}
p({\bf y} | {\bf X}) = \int p({\bf y} | {\bf f}, {\bf X}) p({\bf f} | {\bf X}) \, {\rm d}{\bf f},
\end{equation}
noting that this function space marginal likelihood is weighted by the probability of the data given the function and the probability of the function.  This integral could be evaluated.  However, by noting that a GP is a collection of random variables, where any finite number of which have a joint Gaussian distribution, the marginal probability of our data $\bf y$ is also the joint likelihood of a multivariate normal distribution,
\begin{equation}
p({\bf y} | {\bf X}) = \mathcal{N}(m({\bf X}), {\bf \Sigma}),
\end{equation}
which can be straightforward to evaluate. Thus the marginal likelihood is,
\begin{equation}
p({\bf y} | {\bf X}) = (2 \pi)^{k/2} {\rm det} ({\bf \Sigma})^{-0.5} \exp \left(\frac{-1}{2} ({\bf X} - m({\bf X}))^{T} \, {\bf \Sigma}^{-1} \, ({\bf X} - m({\bf X})) \right),
\end{equation}
which can be evaluated without integrating over all possible function space. While this marginal likelihood expression is clearly more computationally feasible that the integral over functional space is not without it's limitations.  Because it is necessary to calculate the determinant and the inverse of the covariance matrix, typically applied algorithms,  make this a $\mathcal{O}(N^3)$ or $\mathcal{O}(N^2 \log N)$ operation.  This naturally limits the size of the data set for which the likelihood,  and optimisations of the likelihood, can be applied.  

\subsubsection{Making predictions}
If we want to obtain predictive distributions for the output $\bf y_{\star}$ given the inputs $\bf X_{\star}$,  the joint probability distribution of $\bf y$ and $\bf y_{\star}$ is Gaussian and given by
\begin{equation}
p \left( \begin{bmatrix} {\bf y} \\ {\bf y_{\star}} \end{bmatrix} \right) = \mathcal{N} \left( \begin{bmatrix} m({\bf X}) \\ m({\bf X_{\star}}) \end{bmatrix} , \begin{bmatrix} {\bf \Sigma} & {\bf K_{\star} }\\ {\bf K_{\star}}^{T} & {\bf K_{\star \star}} \end{bmatrix}  \right), 
\end{equation}
where the covariance matrices $\bf \Sigma$ and $\bf K$ are computed using the kernel function so that,
\begin{equation}
{\bf \Sigma}_{n, m} = k({\bf X}_{n}, \, {\bf X}_{m}),
\end{equation}
which is an $N \times N$ matrix.
\begin{equation}
{\bf K}_{\star \, n, m} = k({\bf X}_{n}, \, { \bf X}_{\star \, m}),
\end{equation}
which is an $N \times N_{\star}$ matrix, and finally
\begin{equation}
{\bf K}_{\star \star \, n, m} = k({\bf X}_{\star \, n},  {\bf X}_{\star \,m}),
\end{equation}
which is an $N_{\star} \times N_{\star}$ matrix.
The predictions of $\bf y_{\star}$ are again a Gaussian distribution so that,
\begin{equation}
{\bf y}_{\star} \sim \mathcal{N}(\bf \hat{y}_{\star}, \, \bf C),
\label{eq:pred}
\end{equation}
where 
\begin{equation}
{\bf \hat{y}}_{\star} = m({\bf X}_{\star}) + {\bf K}_{\star}^{T} \, {\bf \Sigma}^{-1} \, ({\bf y} - m(\bf X)),
\end{equation}
and 
\begin{equation}
{\bf C} = {\bf K}_{\star \star} - {\bf K}_{\star}^{T} \, {\bf \Sigma}^{-1} \, {\bf K_{\star}}.
\end{equation}

At this point we can make predictions on model properties given a grid of stellar models using equation \ref{eq:pred}.  But these predictions will likely be poor unless we select sensible values for the form and hyperparameters of the mean function and covariance function.  In the following section we detail a number of kernel functions that will be tested against the data.  We will then discuss the method for determining the values of the hyperparameters to be used.

\subsection{Setup of GP Models}\label{sec:setup}

\subsubsection{Tool package}

We adopt a tool package named \textsc{GPyTorch}, which is a GP framework developed by \citet{gardner2018gpytorch}. It is a Gaussian process library based on an open source machine-learning framework PyTorch \footnote{\url{https://pytorch.org}}. The package provides significant GPU acceleration, state-of-the-art implementations of the latest algorithmic advances for scalability and flexibility, and easy integration with deep learning frameworks.\footnote{Source codes and detailed introductions are available on \url{https://gpytorch.ai}.} 
We train GP models on a NVidia Tesla V100 graphics processing unit (GPU) with 32GB GPU Memory. The GPU captivity allows a training dataset with up to $\sim$20,000 data points.

\subsubsection{Training procudure}

The training procedure of a GP model includes training, validating, and testing. 
In the training process, we iteratively optimise hyperparameters of a GP model to learn the underlying function which maps inputs to outputs from on-grid evolutionary tracks (training dataset). In each iteration, the GP model is validated by comparing true and GP predicted values of some off-grid tracks (validating dataset). Although the validating dataset is not directly involved in training hyperparameters, it still constructs the GP model to some extend because the optimal solution is the one that best fits the validating dataset. For this reason, the validating dataset does not give a completely independent validation for a GP model. We hence have a testing process after the training. The testing dataset contents some other off-grid tracks which are reserved  from the training and validating process. The testing dataset are also used to estimate the systematic uncertainties of GP model. 

Here we briefly summary the setup of GP model training. 
We apply an Artificial Neural Network (ANN) including 6 hidden layers and 128 nodes per layer as the mean function. Note that this is not training an ANN to learn the data in detail. The ANN is quickly trained at the beginning to interpret the complex mean function in multiple-dimension space to accelerate the whole training process. In the GP model training, the mean function is normal uninteresting because all the inference effort is spent on estimating the correct covariance function. In our tests, GP models with the linear or the constant mean function could achieve similar results, but it takes more time for models to converge.
The \textsc{Gpytorch} standard likelihood for regression, which assumes a standard homoskedastic noise model, is applied as the likelihood function. We use the negative logarithm of the likelihood as the loss function. The optimiser for training is called `Adam', which is a combination of the advantages of two other extensions of stochastic gradient descent, specifically, Adaptive Gradient Algorithm and Root Mean Square Propagation \citep{kingma2017adam}. More detailed discussions about these choices can be seen in the Appendix \ref{app:A}. 

We set up the so-called `Early Stopping' procedure to decide when to terminate the training. The procedure evaluates GP models on a holdout validation dataset after each iteration. If the performance of the GP model on the validation dataset starts to degrade or stops upgrading after many iterations, then the training process is terminated \citep[see discussions in][]{anzai2012pattern,goodfellow2016deep}. The `Early Stopping' procedure can reduce overfitting and improve the generalisation of GP models. We use a validating error index (defined in Section~\ref{sec:2d})  to monitor the training and terminate it when there is no improvement for 300 iterations.

To save the best learned GP models, we check the validating error index after every iteration. The current model will be saved to replace the last saving if it has the so-far lowest validation errors. This is to say, the final saved model is the one with the best performance in the training process.

\subsubsection{Kernel Function}\label{sec:kernel}

To select the proper kernel function for training GP models, we test four basic kernels and a number of combined kernels. 
The four basic kernels are listed as follow. 
\begin{itemize}
\item {\bf RBF}: Radial Basis Function kernel (also known as squared exponential kernel)
\item {\bf RQ}: Rational Quadratic Kernel (equivalent to adding together many RBF kernels with different lengthscales)
\item {\bf Mat12}: Matern 1/2 kernel (equivalent to the Exponential Kernel)
\item {\bf Mat32}: Matern 3/2 kernel 
\end{itemize}
These four kernels are all universal, and we can integrate each of them against most functions. Every function in its prior has infinitely many derivatives  \citep{williams1996gaussian}. The differences between these kernels, in a simply way, can be understood as their smoothness/flexibility levels. The RBF kernel is very smooth function and can be expressed as a product of a polynomial. It hence suits for the case when the data follow a slowly varying function. The RQ kernel, as a combination of many RBF kernels, is more complex and is able to fit to data with a number of smooth underlying functions (e.g. when the output depends on multiple inputs). On the opposite, the Mat12 gives the absolute exponential kernel, which is hence very spiky. It can fit to any sharp variations in the data. The Mat32 kernel has a smoothness somewhere between the RBF and Mat12 kernels, because it is a combination of an exponential and a polynomial. It is a smooth function but has significantly more extrema than the RBF kernel. For each kernel, there are hyper parameters (e.g. the lengthscale) to modulate the smoothness as well. The hyper parameters can be either prioritised or purely determined in training process. A less smooth kernel like Mat32 with a large lengthscale could behave similarly to the RBF kernel. However, RBF kernel with small lengthscale can not reproduce those extrema in Mat32 kernel.
These differences are not chance coincidence, and the origin of these differences are crucial for interpreting the results. Choosing a good kernel for a particular application is necessary for good predictions. If the kernel function is too spiky for the data, the learnt function could over explain some random variations. On the contrary, if the kernel function is too smooth, it may not fully capture the variability of the underlying function and lead to an increase in bias.
Combining basic kernels can increase the flexibility of a kernel. For instance, the combination of a RBF kernel and a Mat12 kernel is able to fit to both smooth and spiky features in the data. However,  overfitting is a risk when using a combined kernel because the kernel could be over-flexible.

According to the stellar theory, changing fundamental input parameters smoothly changes the dependent outputs. Hence the kernel function for our application needs to be smooth. Moreover, we can notice in Figure \ref{fig:eep} that the observable quantities fast vary at some particular regions (e.g., around the blue hook). This means that a slowly varying function like the RBF kernel may under fit in these areas. We do a number of preliminary studies of training GP models with different kernel functions to choose the proper kernel function. Details of training will be mentioned in Section \ref{sec:2d}. Here we only summary the results. 
We apply four basic kernels and a number of their combinations (RBF + Mat21, RQ + Mat21, Mat32 + Mat21, RBF + Mat32, RQ + Mat32) to train GP models. The combined kernel RBF+Mat21 gives the best fit to the training data, however, its testing errors are large. This indicates that the kernel is too flexible and hence overfits to the data.  
The kernel having the best performance is the Mat32. The GP model with Mat32 fits training data reasonably well and gives the best predictions for the testing models. 
The results match our expectations. What we need is a smooth function but not too smooth to fit to the quick variations at some particular evolutionary phases. The Mat32 kernel is apparently suitable for our application. 

\section{Preliminary Studies}\label{examples}

Before training the whole model grid (with five input dimensions), we start with a number of preliminary studies on low-dimension dataset. 
These preliminary studies are for several purposes. In the 1D problem (training data on a single evolutionary track), we compare GP predictions with the classical interpolator. In the 2D problem, we train GP models on a mass -- {\it EEP} platform to test the performances of using different kernel functions. We also discuss about introducing a new error index for validating and testing GP models instead of using a global error quantity such like Root Mean Square Error.
In the 3D problem, where GP maps three fundamental inputs (mass, EEP, and metallicity) onto observables, we solve the training strategy for the large dataset whose data size excesses the practical limitation. 

\subsection{1D Problem}

\begin{figure}
	\includegraphics[width=1.0\columnwidth]{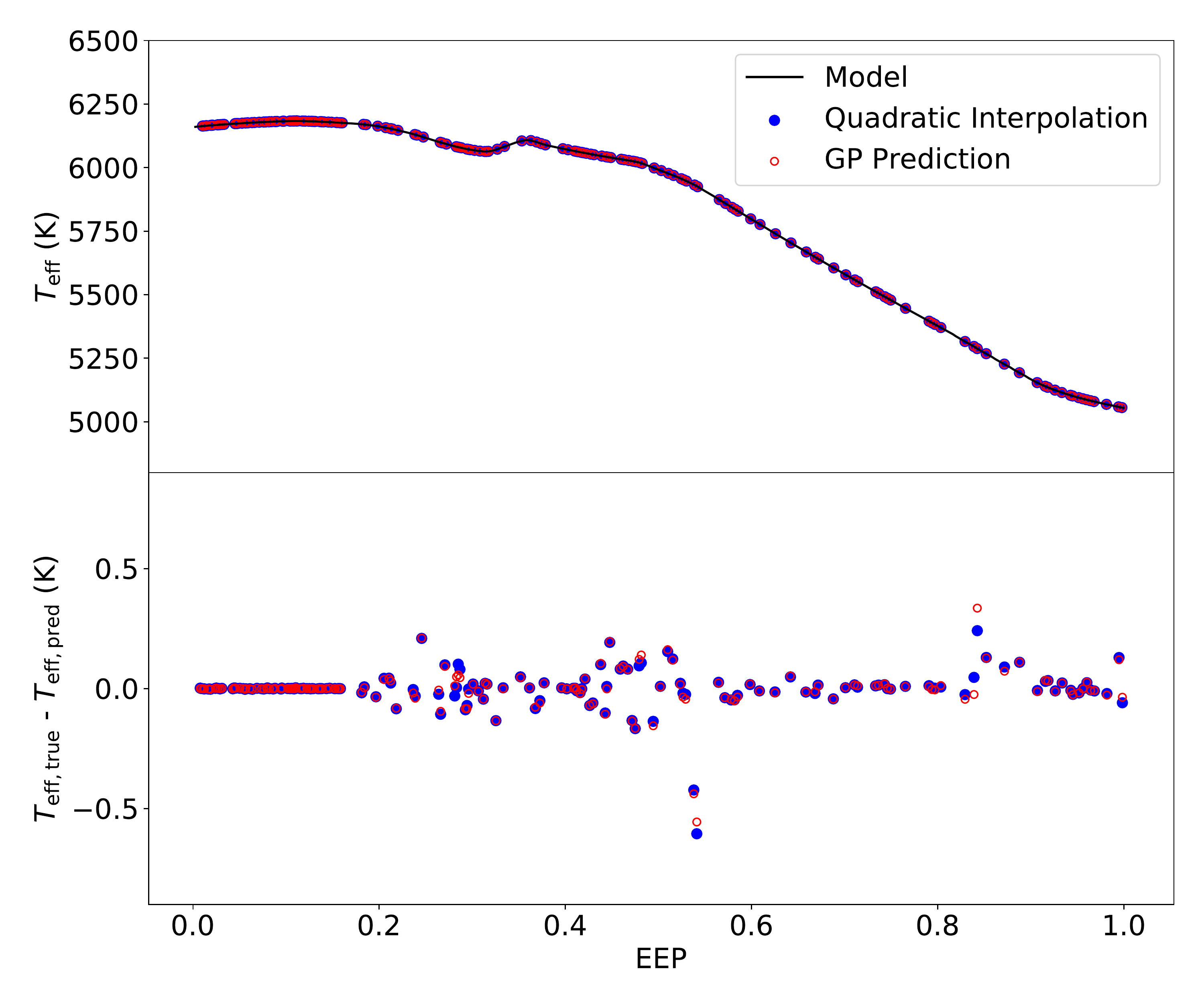}
    \caption{GP application on 1D problem. Models on this track are split into training and testing data by 70-to-30, Top: the evolution of effective temperature for a $\rm 1.1M_{\odot}$ track. The grey line is the evolutionary track computed with \textsc{MESA}; blue and red circles indicate predictions for the testing data from the quadratic interpolator and the GP model. Bottom: residuals of predictions in the top graph. }  
    \label{fig:1dgp}
\end{figure}

We first demonstrate an example of GP application on an 1D problem. We train a GP model using the Mat32 kernel to learn the evolution of effective temperature for a $\rm 1.1M_{\odot}$ track. 
We split the model data points on this track into training and testing data by 70-to-30. We train a GP model which maps  {\it EEP} to effective temperatures and then test GP-predicted effective temperatures with truths. As it can be seen in Figure \ref{fig:1dgp}, the GP model gives very good predictions with residuals less than $\pm$0.5 K. As a comparison, we fit the training data with the quadratic function and use the fitted function to do the same prediction. We find very similar results from the two methods. It suggests that GP can be an alternative of classical interpolators on the 1D problem. 

\subsection{2D Problem}\label{sec:2d}

\begin{figure}
	\includegraphics[width=1.0\columnwidth]{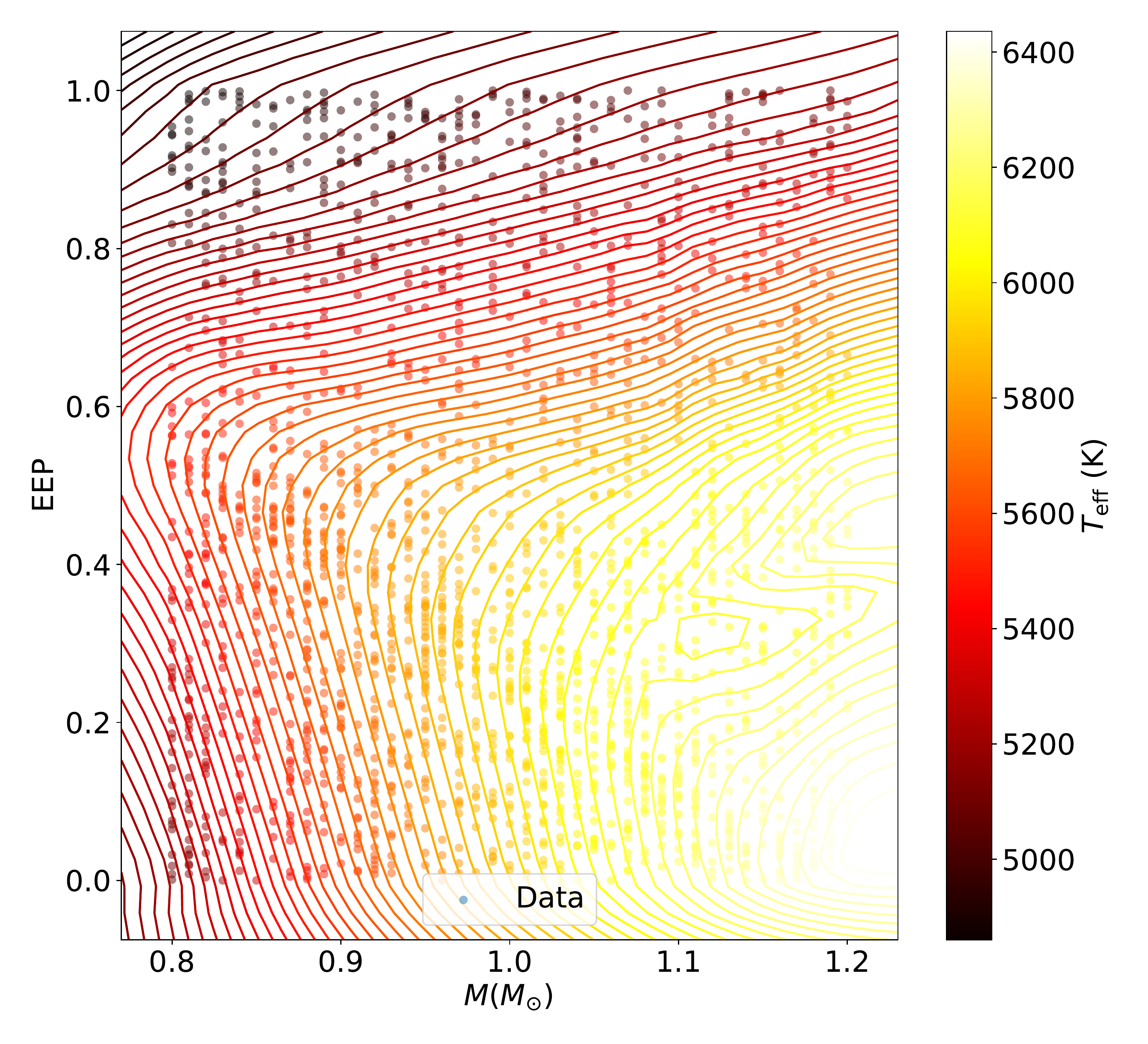}
	\includegraphics[width=1.0\columnwidth]{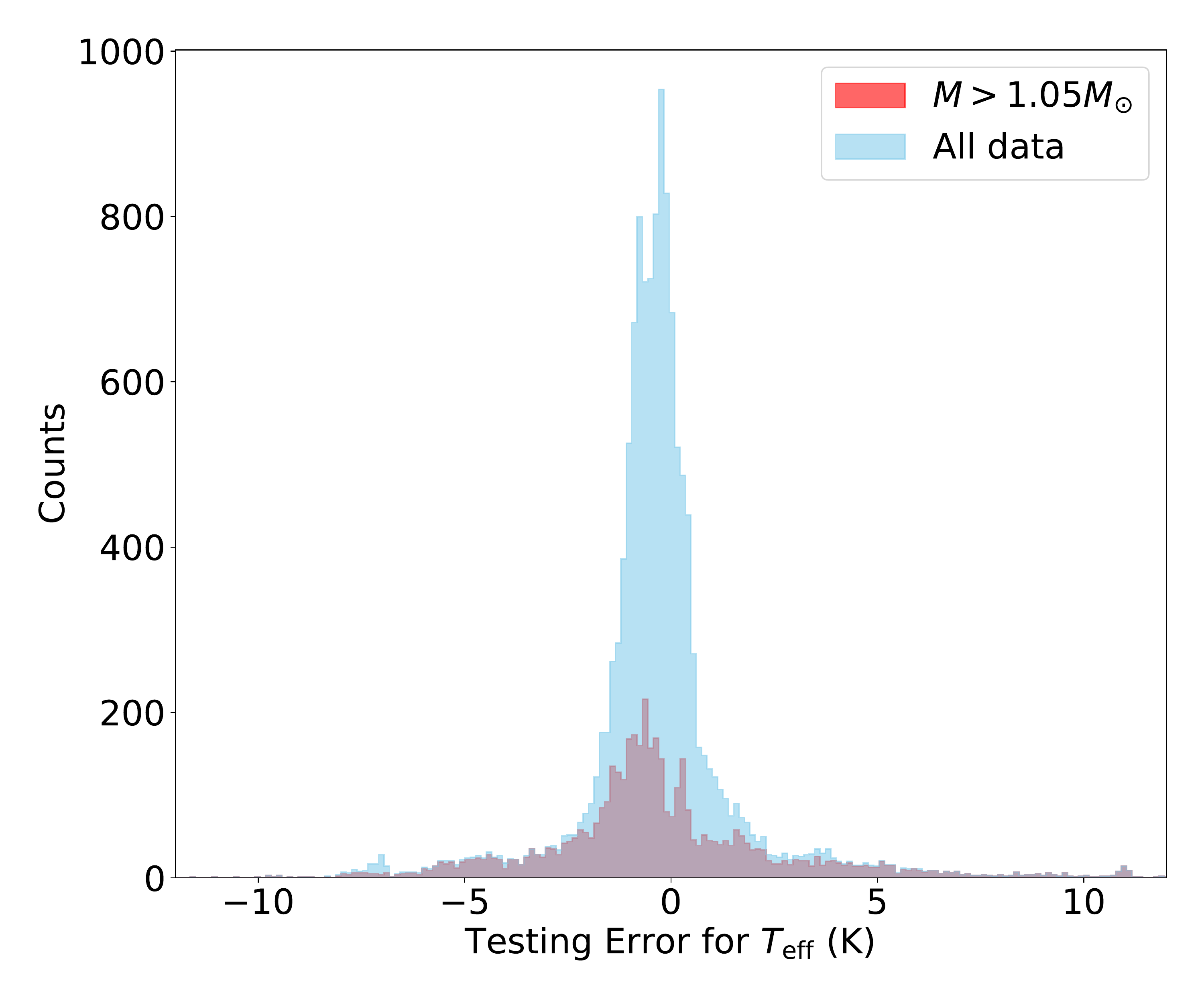}	
    \caption{Top: The 2D GP model for $T_{\rm eff}$. Bottom: probability distributions of validating errors of the GP model. }  
    \label{fig:2dtest}
\end{figure}

As a further step, we train GP models on a 2D problem where GP models map mass and {\it EEP} to the five observable outputs (Outputs = $f(M, EEP)$). Training data are selected from the primary grid with fixed {[Fe/H]}$_{\rm init}$ (0.0), $Y_{\rm init}$ (0.28), and $\alpha_{\rm MLT}$ (2.1). There are 41 evolutionary tracks which content 24,257 models, and we sample 20,000 of them as training data. To validate and test GP models, we compute 44 evolutionary tracks with the same {[Fe/H]}$_{\rm init}$, $Y_{\rm init}$, and $\alpha_{\rm MLT}$ but randomly sampled $M$. We split off-grid tracks half-to-half as validating and testing datasets. 
The script developed based on the \textsc{Simple GP Regression} example \footnote{\url{https://docs.gpytorch.ai/en/stable/examples/01_Exact_GPs/Simple_GP_Regression.html}}. We change the mean function and optimiser in the example and add an early stopping and a model saving modules.  
We follow the aforementioned training procedure to train, validate, and test GP models. We illustrate the learned GP model for effective temperature on the mass--{\it EEP} diagram in Figure \ref{fig:2dtest}. As it shown that, GP transforms the sparse data onto a continuous function and hence is able to predict values for unseen points in the grid. 

It can be seen in Figure \ref{fig:2dtest} that kernels in the area of $M \geq 1.05 {\rm M_{\odot}}$ and ${\it EEP} \leq 0.7$ is more complex than those for other regions. 
There are two regions where global parameters vary relatively fast. The first is around the blue hook and main-sequence-turn-off point ({\it EEP} $\sim$ 0.4) where high-mass tracks sharply turns on the HR diagram. The second is at early subgiant phase ({\it EEP} $\sim$ 0.6) where stars fast restructure. Features in these particular areas are relatively difficult to learn and hence poorly predicted by the GP model.  
When there is a subregion in which the GP model performs worse than other areas, the error distribution would not follow a Normal function. As shown at the bottom of Figure \ref{fig:2dtest}, the density distribution of testing errors form long tails which contents about $10\%$ data. The cases for other two global parameters surface gravity and radius are similar to the effective temperature.

We also find substructures when inspecting testing errors for metallicity and age. The region where the surface metallicity quickly changes is at the early subgiant phase for relatively high-mass tracks. This is because high-mass models maintain shallow convective envelope and hence have strong diffusion effect during the main-sequence stage. At the early subgiant phase, the quick expansion of the surface convective envelope mixes up the settled heavy elements, leading to a fast raise of the surface metallicity. 
The accuracy of age prediction drops down for very old low-mass stellar models. This is because age values vary in a relatively big dynamic range (15 - 50 Gyr) in a small fraction of data points. Poor GP predictions are caused by the low age resolution.

The error distribution causes an issue in validating and testing GP models. What we normally use are some global errors, such as Root Mean Square Error (RMSE), to represent the validating or testing results. For our case, a global error is not able to point out how GP performs in regions where an observable quantity quickly varies. 
We want to have an error index that can reflect the GP model performance in general as well as in those sub-areas. By inspecting the error distributions of all five outputs, we find the data points in the tails (outside the 3 times of full width at half maximum) are around 10\% (8 - 12\% for different outputs).  For the majority (90\%) of data points, which from a Gaussian-like profile, the 68\% confidential interval (1-$\sigma$ uncertainty) can be used to reflect the global accuracy. For the worst 10\% of the data, we could use the 95\% and 99.7\% confidential intervals (2- and 3-$\sigma$ uncertainties) to describe the median and the length of the tail. Thus, we define an Error Index (EI), which is the sum of 68\%, 95\%, and 99.7\% cumulative values of the absolute errors. For the case in Figure \ref{fig:2dtest}, cumulative values at 68\%, 95\%, and 99.7\% are 1.1, 4.9, and 11.1K, which give a testing EI equals to 17.1K. We apply this EI in all following training processes to validate and test GP models. 

We train GP models using different kernel functions to investigate which is the best for our application. We do this with the 2D data because the training is fast to be able to test many different options. As mentioned in Section \ref{sec:kernel}, we find Mat32 is the most suitable kernel for mapping the stellar model grid.

\subsection{3D Problem: Strategy for Large Data Sample}\label{sec:3d}

We apply GP to a 3D problem where GP maps three fundamental inputs, i.e., $M$, {\it EEP}, and {[Fe/H]}$_{\rm init}$ to observables. The main purpose of this preliminary study is investigate the strategy for training large data sample that exceeds the data size limitation of 20,000.
We select training data from the primary grid with $Y_{\rm init}$ = 0.28 and $\alpha_{\rm MLT}$ = 2.1. The training dataset contents $\sim$300,000 data points which is 15 times the limit of training data size (20,000).  For validating and testing purposes, we compute another 174 evolutionary tracks with the same input $Y_{\rm init}$ and $\alpha_{\rm MLT}$ but random input $M$ and {[Fe/H]}$_{\rm init}$. 

We start with sampling 20,000 training data and train a set of GP models. We then obtain testing EI for each output.  For instance, the testing EI for $T_{\rm eff}$ is 23.5K (2.0, 5.8, and 15.7 K at 68th, 95th, and 99.7th). 
We then apply two state-of-the-art approaches designed for large dataset to train the model. These two approaches are named as Stochastic Variational GP (SVGP) and Structured Kernel Interpolation (SKI GP). We only find some minor improvements in the GP predictions. Comparing the testing EI for $T_{\rm eff}$, SVGP gives a results of EI = 24.1K (2.2, 6.8, and 15.1K at 68th, 95th, and 99.7th) and SKI GP ends up with  EI = 22.9K (2.0, 6.1, and 14.8K at 68th, 95th, and 99.7th). Details about these two implementations and discussions about the results can be seen in appendix \ref{app:B}. 

 \begin{figure}
	\includegraphics[width=1.0\columnwidth]{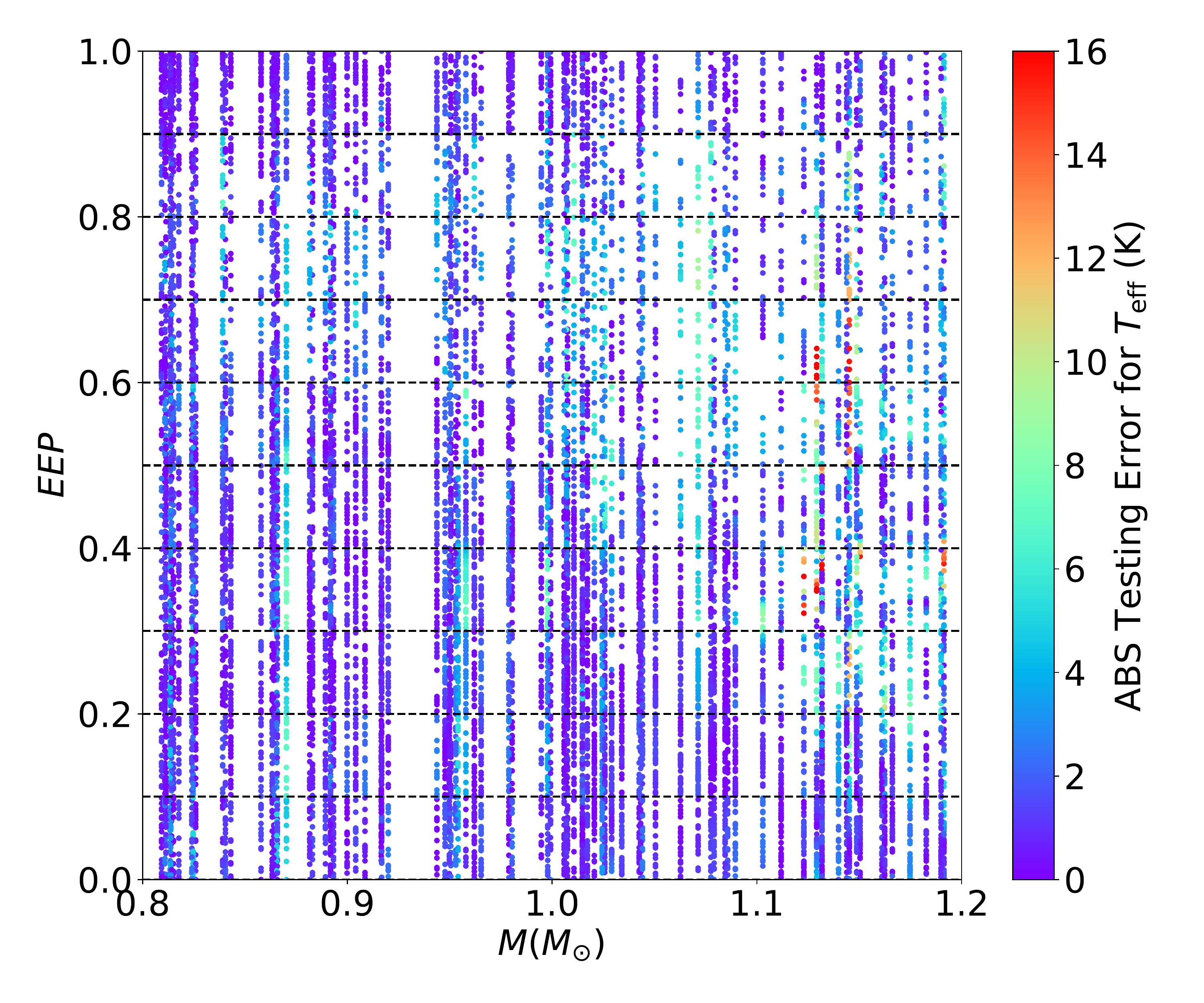}
	\includegraphics[width=1.0\columnwidth]{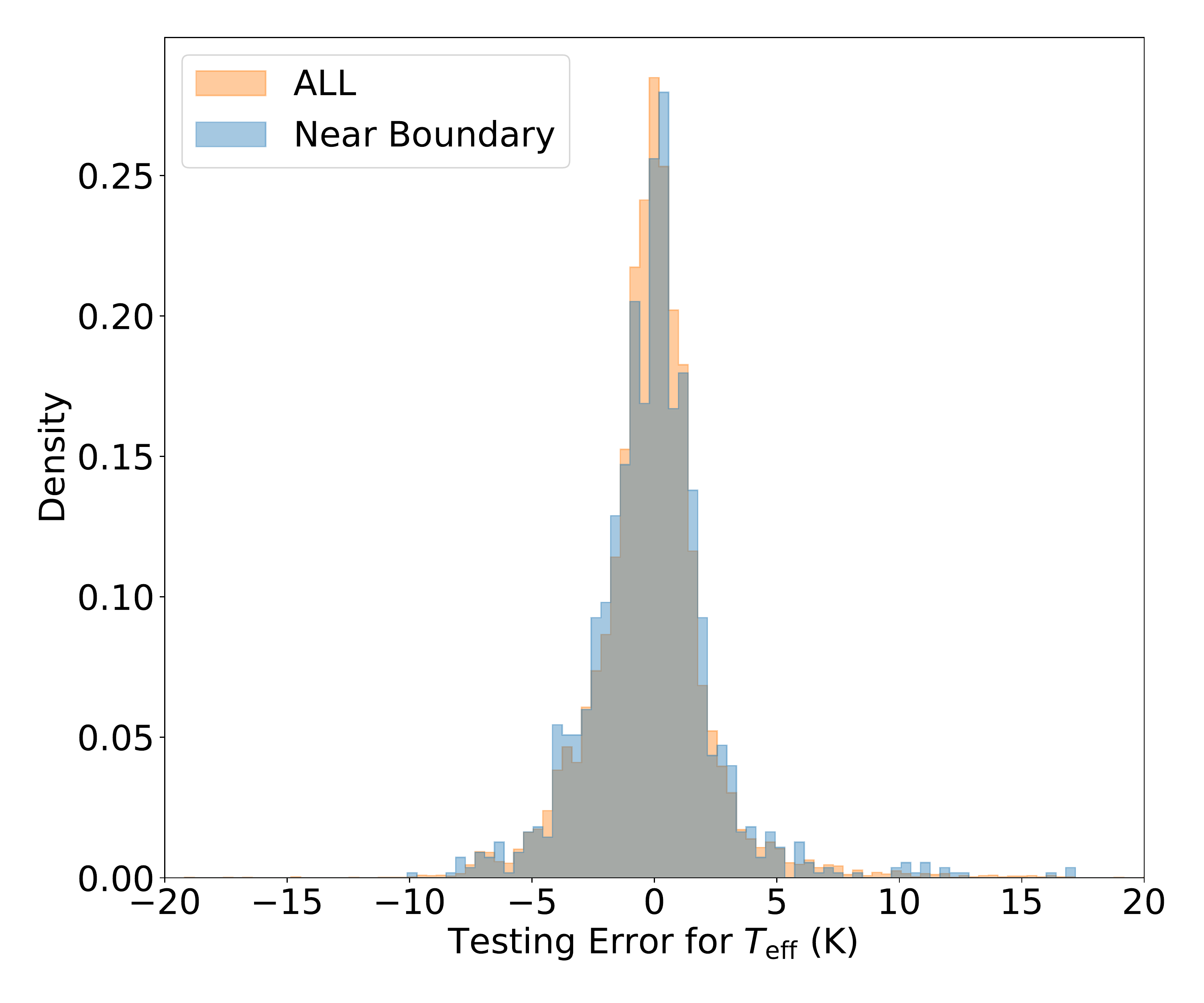}	
    \caption{Top: Testing errors of 3D GP model for $T_{\rm eff}$ on the $M - EEP$ diagram. Dashes indicates section boundaries. Bottom: examination of the edge effects of the section scenario. Probability distributions of testing errors of all testing data and those near the boundary ($\pm$0.01 EEP) in the upper graph are compared.  As it can be seen, testing errors do not raise around the boundary. }  
    \label{fig:3dtest}
\end{figure}

We seek for better strategy for training large data sample.
The GPU memory captivity limits the actual number of data that induce the kernel. This limitation becomes crucial for the high-dimension problems. Because we need much more data given the fact that the parameter space exponentially increases with the dimension. 
A simple way to overcome this issue is breaking the grid into many sections and train GP models for each section separately. 
We divide the training dataset into 10 equal sections by {\it EEP} and sample 20,000 training data in each. A set of GP models are then trained for each {\it EEP} section. 
Using this section scenario, we improve the testing EIs for the five output parameters by around 10\%. For instance, the testing EI for $T_{\rm eff}$ decreases from 23.5 to 21.6K. (1.7, 5.0, 14.9K at 68th, 95th, and 99.7th). EI values for the five outputs before and after sectioning can be seen in Table~\ref{tab:results}.  The section scenario outperforms the SVGP and SKI GP methods. We hence apply it as our training strategy. 
 
The section scenario improves the performance of the GP model, but there is a major concern about the edge effect at the boundary between segments. If a GP model works significantly poorly at these boundaries, it will be difficult to map the systematic errors across the whole parameter space. We hence examine potential edge effects as illustrated in Figure~\ref{fig:3dtest}. We inspect absolute testing errors for each output on the $M-EEP$ diagram. No obvious edge effect is found. We also do a statistical comparison between all errors and those around section boundaries ($\pm0.01EEP$). As shown in the bottom graph, the density distributions of the two samples are very similar to each other.

\section{Augmenting the Stellar Grid}\label{sec:results}

Based on what we find from preliminary studies, we now apply GP to mapping the whole 5D model grid. The setup of GP model is summarised in Table~\ref{tab:setup}. 
The training data are sampled from both primary and additional grids (as described in Table~\ref{tab:grid}). The additional grid increases the grid resolution for relatively high-mass models, and this gives more information about the blue hook for GP to learn.  
We also computed 4,880 off-grid tracks. These off-grid tracks are split by 50-to-50 for validating (in the training progress) and testing (after the training progress) GP models.

The section scenario is applied. For each section, we train a set of GP models for each output parameter with 20,000 training and 20,000 validating data.
The number of sections need to be tested to obtain the best efficiency. To do this, we gradually increase the number of sections from 1 to 100 and track down the changes in testing EI. We find significant improvement from 1 to 10 sections but no further improvements for more than 10 sections. We list the testing EI with different numbers of sections in Table~\ref{tab:results}. It turns out that dividing the grid into 10 sections (corresponding to a 2\% sampling rate) is the most efficient. 

We use GP models for the 10-sections case as our final result. All following analysis and discussion are based on it. %
When testing GP models, we do not section the dataset because the data size limitation for testing is not strict. We sample 100,000 off-grid stellar models as the testing dataset. Note that we do not use models with $\tau \geq$ 20.0Gyr, [Fe/H]$_{\rm surf} \leq$ -0.6dex, or $T_{\rm eff} \geq$ 7000$K$ for testing because we find strong edge effects in those ranges. 

\begin{table*}
	\centering
	\caption{Setup of GP Models}
	\label{tab:setup}
	\begin{tabular}{lcc}
		\hline
		\multicolumn{3}{c}{GP model inputs} \\
		 \hline
		 Parameter & Notation &Range \\ 
                  \hline
                   Mass & $M$ & 0.8--1.2 M$_{\odot}$\\
                   Equivalent evolutionary phase & {\it EEP} & 0 -- 1\\
                   Initial metallicity & [Fe/H]$_{\rm init}$ & -0.5 -- 0.5dex\\
                   Initial helium fraction & $Y_{\rm init}$ & 0.24 -- 0.32 \\ 
                   Mixing-length parameter & $\alpha_{\rm MLT}$& 1.7 -- 2.5\\ 
                   \hline
                   \multicolumn{3}{c}{GP model outputs} \\
                   \hline
                    Parameter & Notation & Trust-worth range$^{a}$ \\ 
                    \hline
                    Effective temperature & $T_{\rm eff}$ & $\leq$ 7000K  \\ 
                    Surface gravity & $\log g$& - \\
                    Radius & $R$ & - \\ 
                    Surface metallicity & [Fe/H] & $\geq$ -0.6dex\\ 
                    Stellar age & $\tau$ & $\leq$ 20Gyr \\ 
		 \hline
		 \multicolumn{3}{c}{Setup of training} \\
		 \hline
		 Item & \multicolumn{2}{c}{Adopted}\\
		 \hline
		 Kernel & \multicolumn{2}{c}{Mat32}\\
		 Mean Function&  \multicolumn{2}{c}{6 layers x 128 notes Neural Network} \\
		  Likelihood Function & \multicolumn{2}{c}{Gaussian Likelihood Function}\\
		 Loss Function & \multicolumn{2}{c}{Exact marginal likelihood}\\
		 Optimiser & \multicolumn{2}{c}{Adam including AMSGRAD variant}\\
		 Termination&  \multicolumn{2}{c}{Early Stoping (monitoring the validating {\it EI}}) \\
		 \hline
		  \multicolumn{3}{c}{$^a$ The ranges without strong edge effects.} \\
	\end{tabular}
\end{table*}

\subsection{Overview of Results}

\begin{figure*}
	\includegraphics[width=2.0\columnwidth]{ 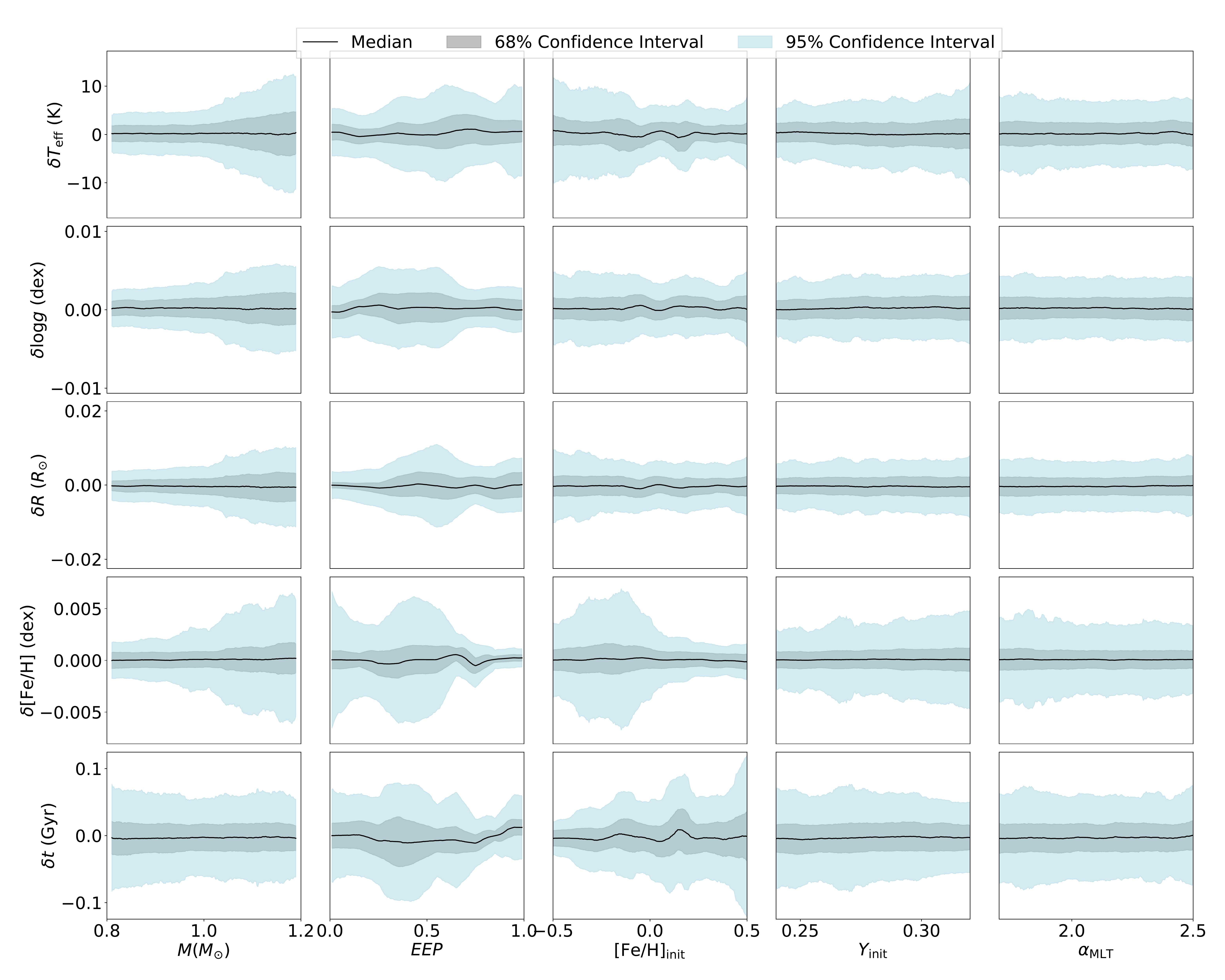}
    \caption{ Roll medians and 68/95\% confidential intervals of testing errors against GP model  inputs. Black solid lines indicate the median value; grey and blue shadowes represent the 68\% and 95\% confidential interval. Testing errors of $T_{\rm eff}$, $\log g$, and $R$ mainly depend on $M$ and{\it EEP}. Metallicity error strongly depends on $M$,{\it EEP}, and [Fe/H]$_{\rm init}$, and age error has a significant correlation to {\it EEP} and [Fe/H]$_{\rm init}$. However, testing errors do not obviously relate to $Y_{\rm init}$ or $\alpha_{\rm MLT}$. } 
  \label{fig:5d_test_vs_input}
\end{figure*}

A overview of testing errors (Truths - GP predictions) can be seen in Figure~\ref{fig:5d_test_vs_input}, where we plot rolling medians and rolling standard deviations for all outputs' errors against fundamental inputs.
Median values are approximate along zero in most plots, indicating good agreement between GP predictions and true values. The 68\% confidence intervals are generally small and their dynamical ranges do not significantly vary across input ranges. 
However, the 95\% confidence intervals have more significant changes and are not well scaled to the 68\% confidence intervals. This corresponds to the tail feature as seen in Figure~\ref{fig:2dtest}. GP predictions are relatively poor in some particular regions. For instance, predictions for the effective temperature are more scattered in high-mass because of the appearances of the blue hook. 
From these results, it can be seen that the model systematic uncertainty is not uniform across the parameter space. Proper estimates of model uncertainty are hence necessary. 


\begin{table*}
	\centering
	\caption{Training and validating errors for GPR Models}
	\label{tab:results}
	\begin{tabular}{cccccccccc}
		\hline
		Model Type&Inputs&$N_{\rm Training}$ &Sampling rate &\multicolumn{5}{c}{Testing Errors (at 68/95/99.7\%)} \\
		 \hline
		 \multicolumn{4}{c}{}& $T_{\rm eff}$ &$\log g$  &$R$  &[Fe/H]$_{\rm surf}$   &$\tau$ \\
		 \multicolumn{4}{c}{}&  (K)& ($10^{-3}$dex) & ($10^{-3}R_{\odot}$)  &  ($10^{-3}$dex)  & ($10^{-2}$Gyr) \\		 
		 \hline
		  GP & 2D & 20,000 x 1 &96\% & 1/5/11 & 1/3/8 & 2/6/14 & 0.5/2/12 &  1/3/9 \\
		 \hline		 
		 GP & 3D & 20,000 x 1 & 5\% & 2/6/16 & 1/4/10 & 3/7/17 &  2/6/22 & 2/7/22 \\
		 GP with 10 sections & 3D & 20,000 x 10 & 50\% & 2/5/15 &1/4/11 & 2/7/17& 1/3/20& 2/6/19\\
		  \hline		 
		GP & 5D & 20,000 x 1 & 0.2\% & 3/9/34 & 2/5/18 & 4/11/36 & 2/7/30 & 3/9/27  \\
		GP with 3 sections & 5D& 20,000 x 3 & 0.6\% &  3/8/27 & 2/5/18 & 3/7/26 & 1/4/24 &3/7/22 \\
		 GP with 5 sections & 5D& 20,000 x 5 & 1\% &  2/7/25 & 1/4/15 & 3/7/24 & 1/4/21 &2/6/22 \\
		 GP with 10 sections & 5D& 20,000 x 10 & 2\% & 2/7/27  & 1/4/14 & 2/7/26 &1/4/20 & 2/6/21\\
		 GP with 20 sections & 5D& 20,000 x 20 & 4\% & 2/7/26  & 1/4/14 & 2/7/27 &1/3/18 & 2/6/22 \\
		  GP with100 sections & 5D& 20,000 x 100 & 20\% & 2/7/25  & 1/4/14 & 2/7/26 &1/3/17 & 2/6/18 \\
		  \hline
	\end{tabular}
\end{table*}

\subsection{Mapping Systematic Uncertainties}\label{sec:sys}

A learned GP model predicts output quantities with uncertainties based on its noise model. In our preliminary studies, uncertainties are properly determined for the 1D and 2D problems, but we find obviously underestimated uncertainties in the 3D problem. In the 5D problem, GP models also predict significantly small uncertainties: they are mostly one order of magnitude smaller than testing errors. This is to say, the learned GP models are over-confident for high-dimension cases. The reason could be the equally spaced training data, from which GP model learn few variations at the scale smaller than the grid step and hence turns to fit with large lengthscale values.   
Because GP models do not give reliable uncertainties, we intend to use testing errors to estimate the systematic uncertainty in GP prediction. As shown in Figure~\ref{fig:5d_test_vs_input}, systematical uncertainties relate to $M$, {\it EEP}, and [Fe/H]$_{\rm init}$ but not to $Y_{\rm init}$ or $\alpha_{\rm MLT}$.  We can treat this as a 3D problem and train another GP model, in which GP model systematic uncertainty is a function of $M$, {\it EEP}, and [Fe/H]$_{\rm init}$. 

\begin{figure}
	\includegraphics[width=1.0\columnwidth]{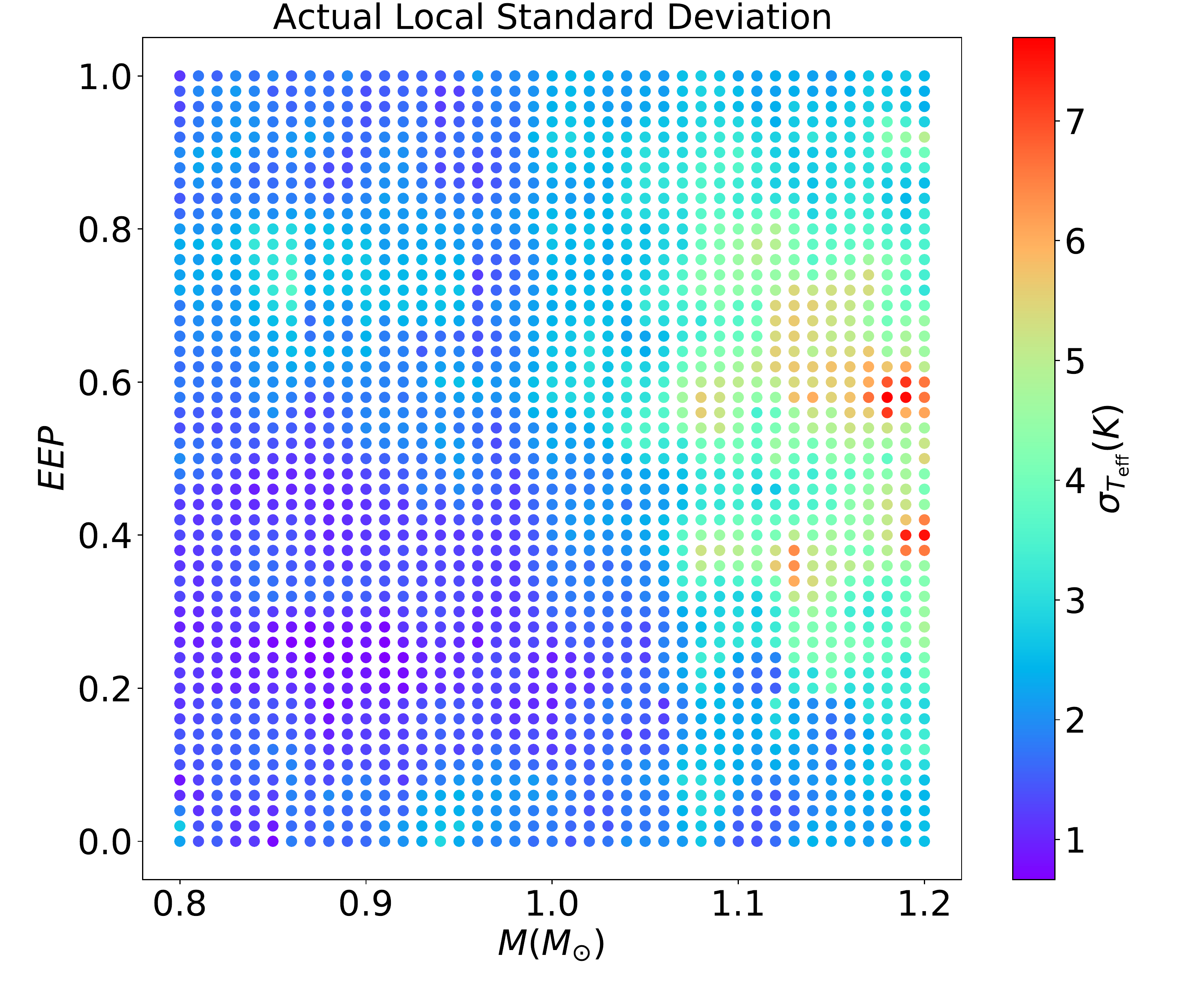}
	\includegraphics[width=1.0\columnwidth]{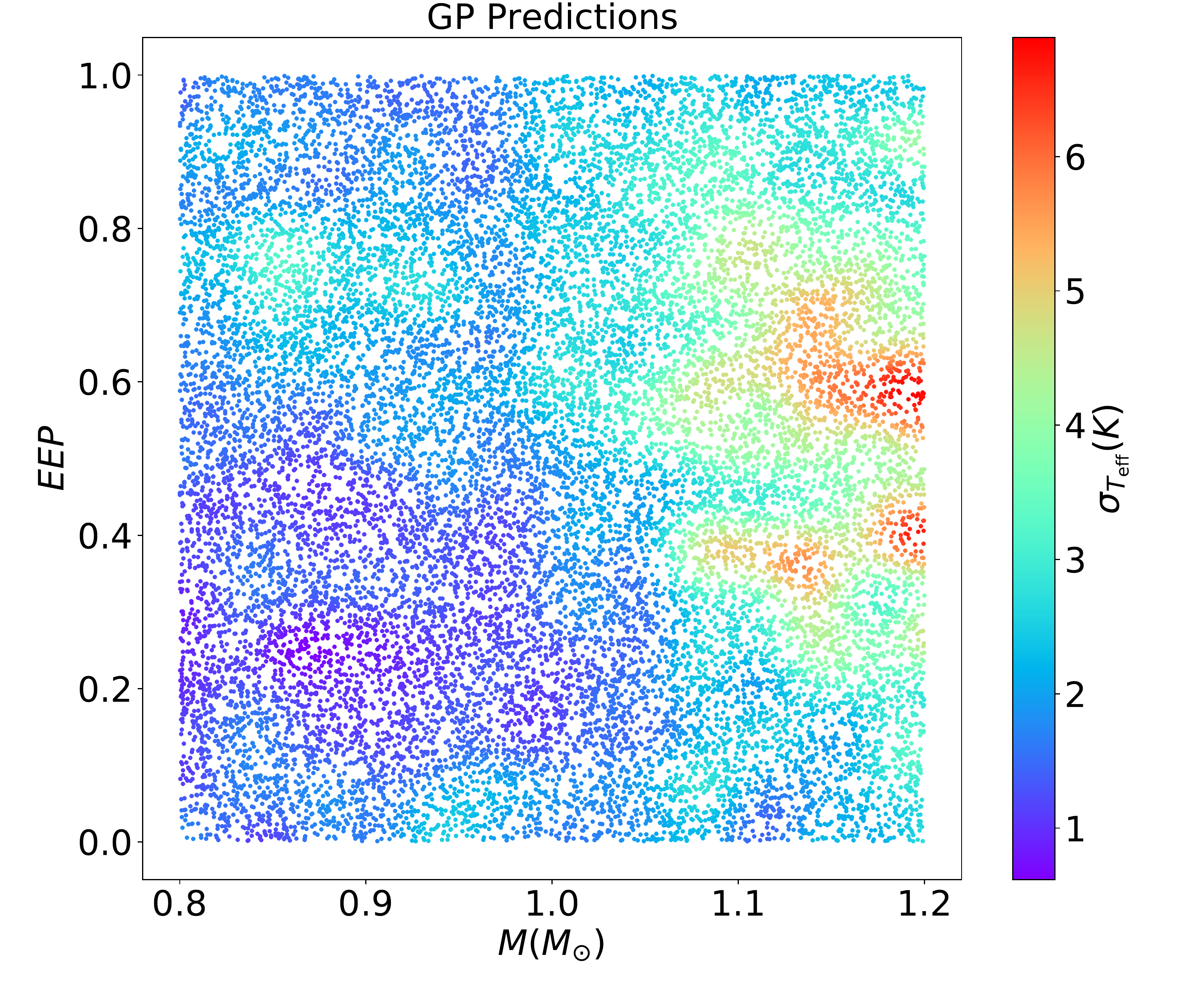}
    \caption{Comparison between actual and GP predicted local systematic uncertainties (1-$\sigma$) for $T_{\rm eff}$ on the $M - EEP$ diagram at [Fe/H]$_{\rm init}$ $\simeq$ 0.0. To calculate the actual local values, we separate the mass range into 40 equally-spaced segments and the {\it EEP} range into 50, and then measure the median testing error for each segment. } 
  \label{fig:5d_sys_teff}
\end{figure}

We inspect the testing errors in the $M$-$EEP$-$\rm [Fe/H]_{\rm init}$ space and find that their local medians vary smoothly. We hence apply the constant mean function and the RBF kernel. The testing dataset contents 100,000 which exceeds the data size limitation. We use the SVGP approach but not the section scenario for this training, because the SVGP can well handle large data following smooth function (see Appendix \ref{app:B} for detailed discussions about SVGP). 
We split the testing error data by 75-to-25 for training and validating.  The variational evidence lower bound (ELBO) is adopted as the loss function because it is designed for when there is too much data for the exact inference. We set up Early Stopping by tracking the RMSE value and terminate training when the RMSE value stops decreasing for 100 iterations. The outputs of GP models are the local medians of testing errors. We use them to infer the systematic uncertainties for five observable quantities (referred as $\sigma_{T_{\rm eff}}$, $\sigma_{\log g}$, $\sigma_{R}$, $\sigma_{\rm [Fe/H]_{\rm surf}}$, and $\sigma_{\tau}$). To differentiate these GP models, we refer to them as GP-SYS models. In Figure \ref{fig:5d_sys_teff}, we compare the actual local systematic uncertainties for $T_{\rm eff}$ at [Fe/H]$_{\rm init}$ $\simeq$ 0.0 with those given by the GP-SYS models. It shows that the GP-SYS model well reproduces the $\sigma_{T_{\rm eff}}$ distributions. 

%

\section{Modelling stars with GP predictions}\label{sec:augmentation}

\subsection{Augmenting the model grid}

\begin{figure*}
	\includegraphics[width=1.3\columnwidth]{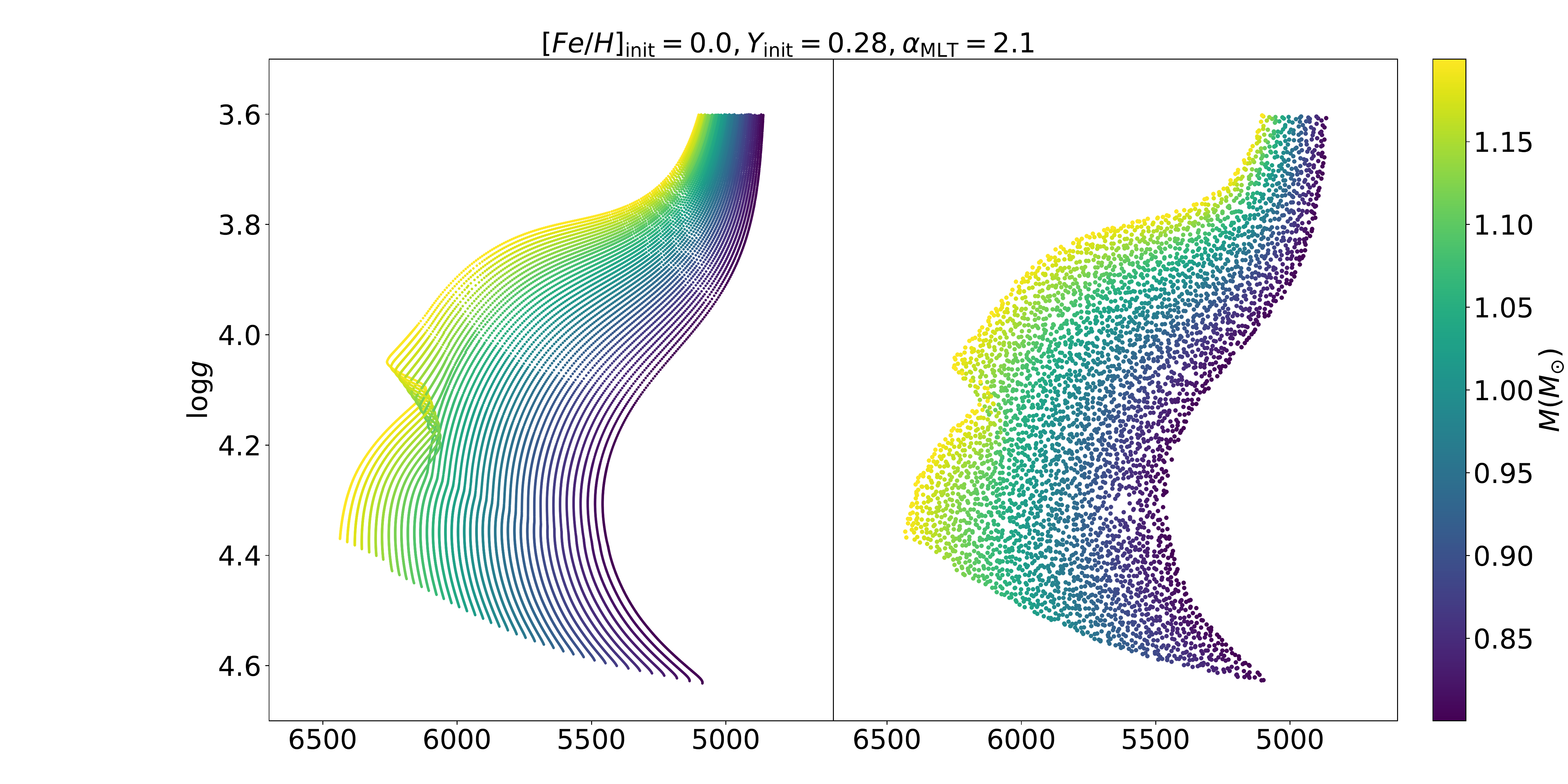}
	\includegraphics[width=0.7\columnwidth]{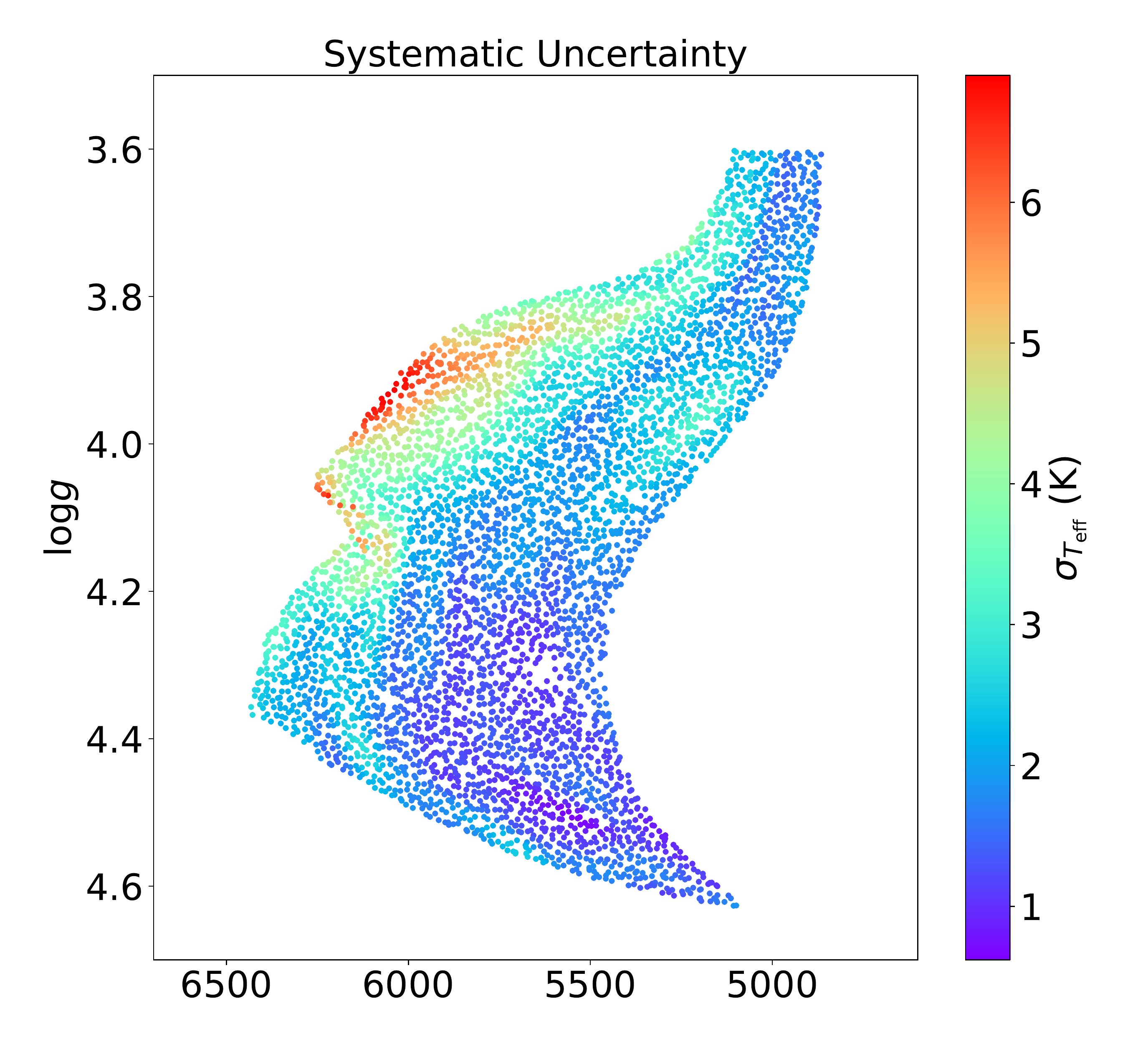}
    \caption{ Left: Comparing the original grid and GP predictions on Kiel diagram. Right: The systematical uncertainties for $T_{\rm eff}$ given by GP-SYS models.} 
  \label{fig:5d_augmentation}
\end{figure*}

Now we are able to use learned GP models to augment the original stellar grid. We randomly sample 5,000,000 data points with uniform distributions for five fundamental inputs ($M$, {\it EEP}, [Fe/H]$_{\rm init}$, $Y_{\rm init}$, and $\alpha_{\rm MLT}$). We then predict output quantities using GP models and their systematic uncertainties using GP-SYS models. This GP-based model dataset can be downloaded following the instruction at \url{https://github.com/litanda/GPGrid}. In Figure~\ref{fig:5d_augmentation}, we demonstrate the original grid, GP predictions, and GP systematic uncertainties on the Kiel diagram. 

\subsection{GP-based Modelling for 1,000 Fake Stars}

As a final test of our method, we use GP-predicted stellar models to characterise 1,000 fake stars to examine whether the method recovers true stellar properties. 
We compute 1,000 fake model stars with the same input physics as the grid but randomly sampled input fundamental parameters. To avoid the edge effect, fake stars are computed in the range of $T_{\rm eff}$ = [4700K, 6800K], $\log g$ = [3.7, 4.6], [Fe/H]$_{\rm surf}$ = [-0.35,0.35], $M$ = [0.85,1.15], {\it EEP} = [0.05,0.95], $Y_{\rm init}$ = [0.25,0.31], and $\alpha_{\rm MLT}$ = [1.8,2.4].
We use four observables, i.e., $T_{\rm eff}$, $\log g$, $R$, and [Fe/H]$_{\rm surf}$, as observed constraints. We apply typical observed uncertainty that is $\pm$50K for $T_{\rm eff}$ (high-resolution spectroscopy), $\pm0.005$dex for $\log g$ (seismology), $3\%$ for $R$ (seismology), and $\pm0.05$dex for [Fe/H]$_{\rm surf}$ (high-resolution spectroscopy). Observed value for each constraint is calculated with true value plus a random noise which follows a Gaussian distribution.  

We fit fake stars using the Maximum Likelihood Estimate (MLE) method. Note that the variance term in the MLE function contents observed uncertinaty and also the model systematic uncertainty determined with GP-SYS models ($\sigma^{2} = \sigma_{obs}^{2} +  \sigma_{sys}^{2} $). We measure the 16th, 50th, and 84th percentiles of the likelihood distribution to estimate a parameter and its uncertainty.

\begin{figure*}
	\includegraphics[width=1.9\columnwidth]{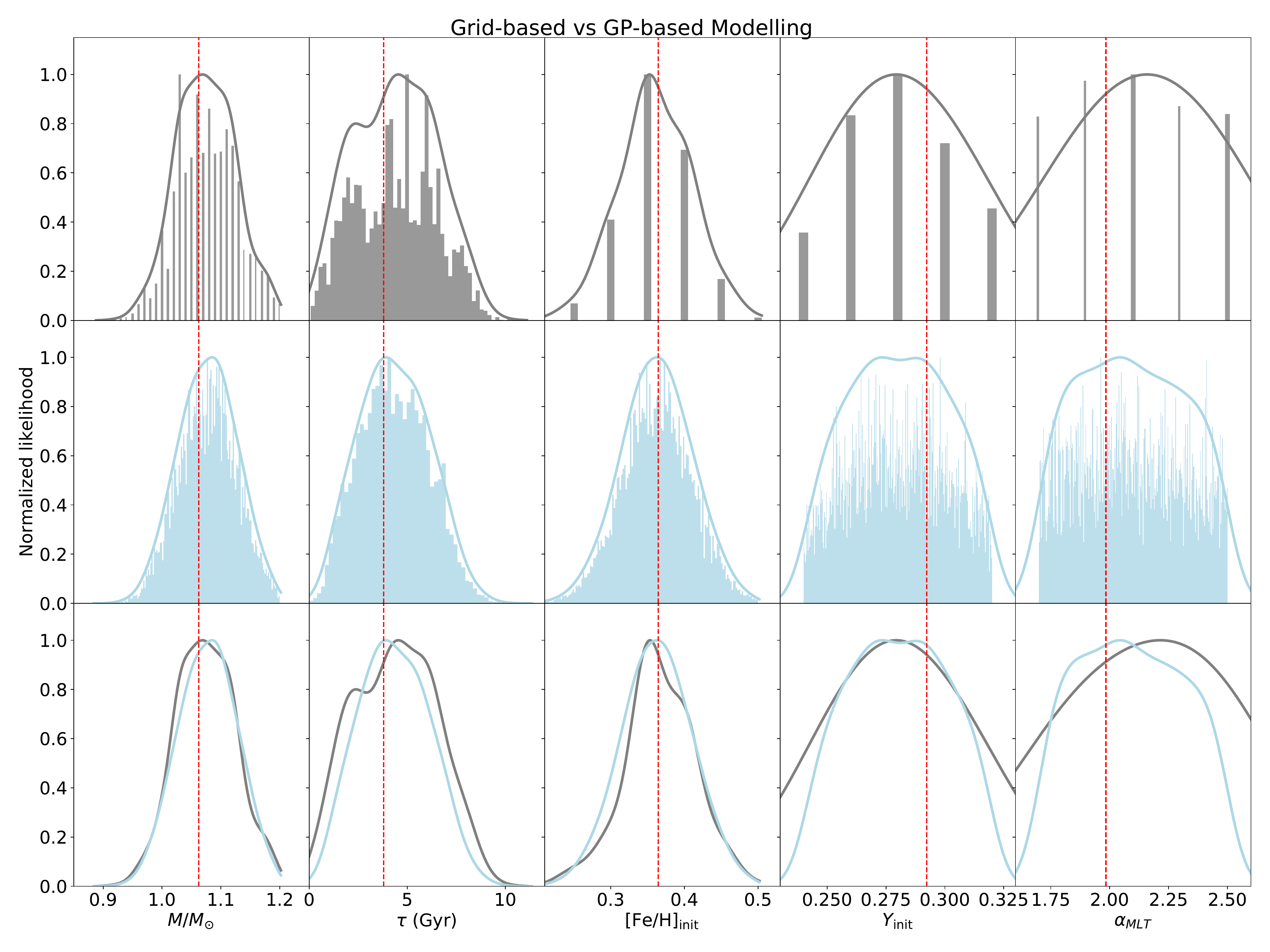}
    \caption{Probability distributions of five fundamental parameters from grid-based (top row) and GP-based modelling (middle row) for a fake star. Grey solid lines in the top row and blue solid lines in the middle row are the kernel density of probability distributions. The bottom row demonstrates comparisons between kernel densities based on the two methods. True fundamental parameters, indicated by red dashed lines, are $M$ = 1.062$\rm M_{\odot}$, $\tau$ = 3.79Gyr, $\rm [Fe/H]_{init}$ = 0.364, $Y_{\rm init}$ = 0.292, and $\alpha_{\rm MLT}$ = 1.984. Observed constraints for this fake star are $T_{\rm eff}$ = 5652$\pm$50K, $\log g$ = 4.424$\pm$0.005, [Fe/H]$_{\rm surf}$ =  0.31$\pm$0.05, and $R$ =  1.047$\pm$0.031$R_{\odot}$. } 
  \label{fig:fit_comparison}
\end{figure*}

We present inferred stellar parameters for a representative fake star in Figure \ref{fig:fit_comparison}. Observed constraints for this fake star are $T_{\rm eff}$ = 5652$\pm$50K, $\log g$ = 4.424$\pm$0.005, [Fe/H]$_{\rm surf}$ =  0.31$\pm$0.05, and $R$ =  1.047$\pm$0.031$R_{\odot}$. True fundamental parameters are $M$ = 1.062$\rm M_{\odot}$, $\tau$ = 3.79Gyr, $\rm [Fe/H]_{init}$ = 0.364, $Y_{\rm init}$ = 0.292, and $\alpha_{\rm MLT}$ = 1.984 (red dashes). We fit with both the original model grid and the GP predictions for comparing.
As it shown that the GP-based modelling has a completed statistical sampling and hence gives more sensible posterior distributions than the grid-based modelling. The improvement for the age is obvious. GP-based modelling infers an age of $4.3^{+2.0}_{-1.8}$Gyr, which is more precise than that determined with the original grid ($4.4^{+2.2}_{-2.4}$Gyr). The age estimate with the original grid does not actually converge because of under sampling. For initial metallicity, initial helium fraction, and the mixing-length parameter, GP-based models make it possible to properly estimate them without computing a very fine model grid. Thus, GP is an efficient tool to augment a typical stellar model grid and overcome the under sampling issue, as a result, improves the precision of estimate.  

\begin{figure*}
	\includegraphics[width=1.8\columnwidth]{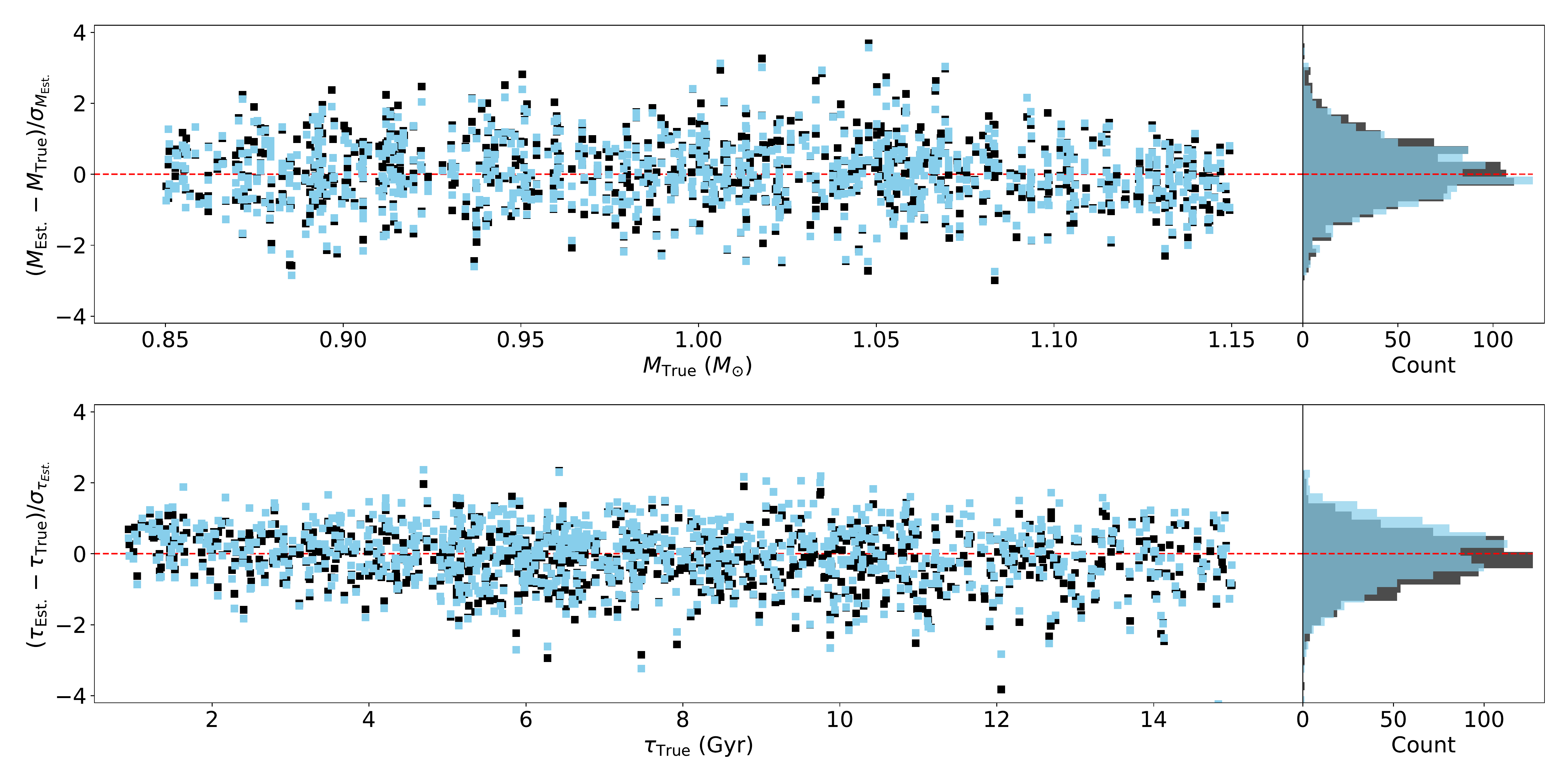}
    \caption{Differences between true and estimated stellar masses and ages over their estimated uncertainty of 1,000 fake stars. Black and blue symbols represent inferences with the original grid and with GP predictions. Count distributions of offsets are demonstrated on the right side.} 
  \label{fig:fake_test}
\end{figure*}

We compared estimates for the 1,000 fake stars and find remarkable improvements in age estimates. 
The average precisions are 4.7\% for mass and 26\% for age with the original grid and 4.5\% and 19\% with GP predictions. 
The accuracy of GP-based modelling is examined as illustrated in Figure \ref{fig:fake_test}, in which we compare modelling solutions with fake stars'  true masses and ages.
Good consistence is found, and we also see that differences nicely follow a normal distribution, saying that the fitting is only affected by random noises in observations. 
Comparing between grid-based and GP-based modelling, their results for the mass both consist with truths, but the median age difference shifts to around -0.2 for the grid-based modelling. The underestimations of age is due to the in-completed sampling in the original grid.

\section{Conclusions}\label{sec:conclusion}

In this work, we apply a machine-learning algorithm that involves a Gaussian process as an interpolator to augment a stellar grid. We train GP models to convert a sparse model grid into continuous functions, which map five fundamental inputs to observable quantities. We find good precision and accuracy in GP-based interpolations when testing them with off-grid models. A GP-predicted model dataset is then generated and we use it to model 1,000 fake stars. Comparing with the original sparse grid, GP-predicted models have complete sampling across the parameter space and hence improve the accuracy and the precision of estimates, particularly for the stellar age. Moreover, the archived continuous functions make it possible to do statistical analysis sampling like Markov Chain Monte Carlo, which provide statistically-sound estimates for stellar parameters.

The application of GP's make it efficient to generate a statistically-sound model dataset based on a sparse model grid. It works well for high-dimension case (up to 5D in this work) and this can be a big advantage compared with traditional interpolators. The choice of kernel is crucial for the success of a particular case. We hence do a number of preliminary studies for the best option. A limitation of the training step for the GP is the size of the training data set and the requirement to perform matrix inversion and calculate the matrix determinant. For training a normal GP model, the practical training data size upper limit is around 20,000 (the number depends on the capacity of GPU device) which is apparently not big enough for the high-dimension grid including $\sim$10,000,000 models. To overcome this issue, we section the training data according to their evolutionary stage ({\it EEP}) and train GP models for each section. This section scenario significantly improves overall accuracy of GP predictions and we find no edge effects at the boundary of sections. When inspecting the systematic uncertainty of GP models, we notice that GP predictions are very over-confident in the high-dimension problem: the uncertainties given by GP models are mostly smaller than testing errors by an order of magnitude. Because of this,  we use the testing errors to estimate systematic uncertainties across the parameter space. Eventually, we provide a GP-based model dataset including 5,000,000 models with randomly sampled fundamental inputs.

We use GP-based models to characterise 1,000 fake stars to examine whether truths of stellar properties can be recovered. We find that GP-based masses and ages are consistent with the injected truth values. The uncertainties are dominated by observational noise, saying that, the systematic uncertainty due to the GP approximation does not obviously affect the modelling on interferences. Comparing with the probability distributions of original sparse grid, GP models are fully sampled in the input range and hence improve the accuracy and precision of inferred parameters. The improvement is remarkable for the precision of stellar age (by 7\%). Moreover, the continuous sampling makes it possible to properly estimate some fundamental parameters which are sparse in the grid, e.g., the helium fraction. These results indicate that the method demonstrated in this work is reliable and efficient for interpolating an established model grid and it can improve the modelling solutions because of the statistically-sound sampling. 

\section*{Acknowledgements}
This work has received funding from the European Research Council (ERC) under the European Union’s Horizon 2020 research and innovation programme (CartographY GA. 804752). R.A.G. Acknowledges funding from the PLATO CNES grant.
Development of \textsc{GPyTorch} is supported by funding from the Bill and Melinda Gates Foundation, the National Science Foundation, and SAP. We also thank Dr. Victor Aguirre B{\o}rsen-Koch for advices regarding this work.


\section*{Data availability}

The data underlying this article are available in Zenodo, at \url{https://doi.org/10.5281/zenodo.5879255}. 



\bibliographystyle{mnras}
\bibliography{ref} 

\appendix{}

\section{Setup of GP model training}\label{app:A}

This section includes detailed discussions about the selections of mean function, likelihood and loss function, and optimiser. 

\subsection{Mean Function}

We first investigate the mean function. As discussed above, the data distribution is generally smooth but complex in some regions of parameter space (e.g., the sub-giant hook).  Although the choice of mean function is not crucial for training GP models, we find that using a constant or a linear mean function leads to a significantly long training time. Hence, we apply a neural network mean function which is flexible enough to manage both simple and complex features to accelerate the training. We adopt an architecture based on that of  \citet{2021MNRAS.tmp.1343L} comprising 6 hidden layers and 128 nodes per layer. All layers are fully-connected and the output of each layer, except for the last, is transformed by the Exponential Linear Unit (ELU) activation function \citep{2015arXiv151107289C}. The ELU activation function provides a smooth function from inputs to outputs, which is preferred over its more common, faster counterpart, the Rectified Linear Unit (RELU).

\subsection{Likelihood and Loss Function}

Our training dataset is a theoretical model grid,  hence there is no observed uncertainty for each data point, but a tiny random uncertainty exists due to the approximations in the \textsc{MESA} numerical method. We model this using $\sigma_{w}$.  This noise model is then assumed to be a Gaussian function with a very small variance.  
A likelihood specifies the mapping from input values $f(X)$ to observed labels $y$.
We adopt the the standard likelihood for regression which assumes a standard homoskedastic noise model whose conditional distribution is

\begin{equation}\label{eq:likelihood}
p(y|f(x)) = f + \epsilon, \epsilon \sim \mathcal{N}(0, \sigma^{2}),
\end{equation}
where $\sigma$ is a noise parameter. 
We used a small and fixed noise parameter and run a few tests.  However, the strict noise parameter makes GP models hard to converge. When this noise parameter is set as free, it reduces to a small value anyway in the training progress because it is data-driven.  For this reason, we did not put strict constraint for or prioritise this noise parameter.  In practice, we only set up a loose upper limit ($\sigma$  < 0.1) to speed up the training. One thing should be noted that a GP model with a large noise parameter is not a proper description for the stellar grid. Because of this, we only adopt GP models with $\sigma$ $\lesssim 10^{-4}$. We train GP models using the negative logarithm of the likelihood function as the loss function.  

\subsection{Optimiser}

We compared two optimisers named SGD and Adam. Here SGD refers to Stochastic Gradient Descent, and Adam is a combination of the advantages of two other extensions of stochastic gradient descent, specifically, Adaptive Gradient Algorithm and Root Mean Square Propagation. 
The SGD optimiser in the \textsc{GPyTroch} package involves the formula given by \citet{sutskever2013importance}. The formula makes it possible to train using stochastic gradient descent with momentum thanks to a well-designed random initialisation and a particular type of slowly increasing schedule for the momentum parameter. The application of momentum in SGD could improve its efficiency and make it less likely to stuck in local minimums. On the other hand, the Adam optimiser includes the 'AMSGrad' variant developed by \citet{47409} to improve its weakness in the convergence to an optimal solution. With these new developments, the two optimisers give very similar results. We finally choose Adam because it works relatively efficiently and stable.  
We adaptive learning rate in the training process. Our training starts with a learning rate of 0.01 and decreases by a factor of 2 when the loss value does not reduce in previous 100 iterations.

\section{State-of-the-art implementations for large dataset}\label{app:B}

In section \ref{sec:3d}, we investigate the strategy for large dataset. We test two State-of-the-art approaches that designed for training big data whose data size is more than the limit of a GP model. Here we mention some details about the two methods and the results. 

We first consider the Stochastic Variational GP (SVGP) approach based on the \textsc{GPyTorch ApproximateGP} module. 
We train our data based on the SVGP example on \url{https://docs.gpytorch.ai/en/v1.1.1/examples/04_Variational_and_Approximate_GPs/SVGP_Regression_CUDA.html}. 
SVGP is an approximate scheme rely on the use of a series of inducing points which can be selected in the parameter space. It trains using mini batches and hence is able to deal with large data size. The other key point of SVGP is the number of inducing points. Because the kernel is only built on these points, the number determines the complexity of kernel. When the underlying function is simple, for instance, a power law, a small number of inducing points  is enough. For our application, a large number of inducing points are required. Underlying principles and detailed descriptions of this approach can be found in \citet{hensman2015scalable}. In our tests on 3D problem, we find a practical issue with the SVGP approach. When we load in a large training sample which takes a lot GPU memory, the rest memory can capacitate only 10,000 inducing points. This is to say, we use more training data but sacrifice the kernel complexity.  The result shows that using SVGP model does not improve the GP predictions compared with normal GP model. For instance, the 68th, 95th, and 99.7th testing errors for $T_{\rm eff}$ are 2.2, 6.8, and 15.1K (EI = 24.1 K) for the SVGP and 2.0, 5.8, and 15.7 K (EI = 23.5K) for the normal GP model. This is because the evolutionary feature are complex across multiple dimensions. Reducing the kernel complexity is not ideal. We conclude that the SVGP is suitable for training large data which have relatively simple variations but not a good choice for training the model grid.  

We also investigate another approach designed for large dataset named Structured Kernel Interpolation (SKI GP). SKI GP was introduced by \citet{wilson2015kernel}. It produces kernel approximations for fast computations through kernel interpolation and is a great way to scale a GP up to very large datasets (100,000+ data points).
We follow the example on \url{https://docs.gpytorch.ai/en/stable/examples/02_Scalable_Exact_GPs/KISSGP_Regression.html} to develop our script. 
We run a few tests to train a 3D SKI GP model with 100, 000 training data. Compare with the Normal GP and SVGP, its testing errors for $T_{\rm eff}$ are slightly improved to 2.0, 6.1, and 14.8K (EI = 22.9K). However, the further test on the 5-dimension data is not ideal: a SKI GP model using 100, 000 training data performs much worse than a normal model with only 20,000 training data. The poor behaviour consists with what has been discussed by \citet{wilson2015kernel}: the SKI GP poorly scale to data with high dimensions, since the cost of creating the grid grows exponentially in the amount of data. We attempt to make some additional approximations with the \textsc{GpyTorch AdditiveStructureKernel} module. It makes the base kernel to act as one-dimension kernels on each data dimension and the final kernel matrix will be a sum of these 1D kernel matrices. However, the testing errors are not significantly improved. 







\bsp	
\label{lastpage}
\end{document}